\documentclass{aa} 
\usepackage{graphicx}
\usepackage{mathtools}
\usepackage{bclogo}
\usepackage{txfonts}
\usepackage[thinc]{esdiff}
\usepackage{hyperref}
\hypersetup{colorlinks,citecolor=blue,linkcolor=blue}
\begin{document} 
   \title{Effect of stellar rotation on the development of post-shock instabilities during core-collapse supernovae}
   \titlerunning{Stellar rotation and post-shock instabilities during core-collapse}

   \author{A.-C. Buellet$^{1}$, T. Foglizzo$^{1}$, J. Guilet$^{1}$, E. Abdikamalov$^{2}$}
   \authorrunning{A.-C. Buellet, T. Foglizzo, J. Guilet, E. Abdikamalov}

   \institute{$^{1}$ Université Paris-Saclay, Université Paris Cité, CEA, CNRS, AIM, 91191, Gif-sur-Yvette, France\\
   $^{2}$Department of Physics and Energetic Cosmos Laboratory, Nazarbayev University, Astana 010000, Kazakhstan\\
              \email{anne-cecile.buellet@cea.fr}}

   \date{Received  25 December 2022; accepted 11 April 2023}

% \abstract{}{}{}{}{} 
% 5 {} token are mandatory
 
  \abstract
    {The growth of hydrodynamical instabilities is key to trigger a core-collapse supernova explosion during the phase of stalled accretion shock, immediately after the birth of a proto-neutron star (PNS). Stellar rotation is known to affect the standing accretion shock instability (SASI) even for small rotation rates, but its effect on the onset of neutrino-driven convection is still poorly known.}
  {We assess the effect of stellar rotation on SASI when neutrino heating is taken into account, and the effect of rotation on neutrino-driven convection. The interplay of rotation with these two instabilities affects the frequency of the mode $m=2$ which can be detected with gravitational waves at the onset of a supernova explosion. }
   {We use a linear stability analysis to study the dynamics of the accreting gas in the equatorial plane between the surface of the PNS and the stationary shock. We explore rotation effects on the relative strength of SASI and convection by considering a large range of specific angular momenta and neutrino luminosities.} 
  {The nature of the dominant non-axisymmetric instability developing in the equatorial postshock region depends on both the convection parameter $\chi$ and the rotation rate. Equatorial convective modes with $\chi\gtrsim 5$ are hampered by differential rotation. At smaller $\chi$, however, mixed SASI-convective modes with a large angular scale $m=1,2,3$ can take advantage of rotation and become dominant for relatively low rotation rates at which centrifugal effects are small.
  For rotation rates exceeding $\sim30\%$ of the Keplerian rotation at the PNS surface, 
  a new instability regime is characterised by a frequency which,
  when measured in units of the postshock velocity and radius 
  $v_{\rm sh}/r_{\rm sh}$, is nearly independent of the convection parameter $\chi$. A strong prograde $m=2$ spiral dominates over
  %the growth rate of the dominant mode depends weakly on the rate of neutrino heating which highlights the arising of a new instability regime. Similarly, the frequency of this mode is surprisingly independent of the heating rate, with a strong prograde spiral $m=2$ dominating over 
  a large parameter range favourable to the production of gravitational waves. 
  In this regime, a simple linear relation exists between the oscillation frequency of the dominant mode and the specific angular momentum of the accreted gas.} 
   {Three different regimes of postshock instabilities can be distinguished depending on the rotation rate. For small rotation rates (less than $10\%$ of the Keplerian rotation at the PNS surface), differential rotation has a linear destabilising effect on SASI and a quadratic stabilising or destabilising effect on the purely convective equatorial modes depending on their azimuthal wavenumber. Intermediate rotation rates ($10$ to $30\%$ of the Keplerian rotation) lead to the emergence of mixed SASI/convection/rotation modes involving large angular scales. Finally, strong rotation erases the influence of the buoyancy and heating rate on the instability. This independency allows for a reduction of the parameter space, which can be helpful for gravitational wave analysis.}

   \maketitle
%
%-------------------------------------------------------------------

\section{Introduction}

The death of massive stars begins with the collapse of their iron core, which forms a proto-neutron star (PNS) for zero-age main sequence stellar masses in the range [10, 60]~$\rm M_\odot$ \citep{Woosley+2002,woosley17}. The bounce creates a shock wave that propagates outward, gradually losing energy, dissociating iron atoms until it stalls. The development of multidimensional instabilities during the phase of stalled shock impacts both the revival of the shock \citep[e.g.][]{Herant+1994,JankaMuller1996,CouchOConnor2014,Takiwaki+2016} and the multimessenger signature \citep[e.g.][]{Tamborra+2013,Janka+2016,Kuroda+2016,Muller2020,BurrowsVartanyan2021}. Among them, the stationary accretion shock instability (SASI) \citep{Blondin+2003,BlondinMezzacappa2006} can generate shock oscillations and contributes to push the shock further up \citep{Scheck+2008,MarekJanka2009,Hanke+2013}. The magnitude of this effect depends on the concurrent growth of the neutrino-driven convection, which can generate turbulence in the post-shock region \citep{Abdikamalov+2015,Radice+2016}. Different paths to explosions dominated either by neutrino-driven convection or SASI may occur depending on the precise physical conditions determined by the progenitor structure \citep{Muller+2012,Murphy+2013,Fernandez+2014} and the magnitude of pre-collapse turbulence asymmetries \citep{CouchOtt2013,Muller+2017}. 

The development of neutrino-driven convection and/or SASI leaves clear signatures in gravitational waves \citep{Murphy+2009,Kuroda+2016,andresen17}. The eigenfrequencies deduced from a perturbative analysis can be recognised in the gravitational wave signal \citep{TorresForne+2018,TorresForne+2019} and can be used to constrain parameters such as the PNS mass and radius or the shock radius \citep{TorresForne+2019universal, SotaniTakiwaki2020, Sotani+2021}. These asteroseismic properties can also help constrain the equation of state at nuclear densities \citep{Kuroda+2016,Sotani+2017,Kuroda+2022}.

The neutrino signal is also a precious messenger carrying direct information on the frequency of large-scale shock oscillations induced by SASI, currently detectable for a galactic supernova \citep{Tamborra+2013}. In addition, the correlated detection of neutrinos modulation and gravitational wave could constrain the instability at work \citep{Kuroda+2017,Shibagaki+2021}.

A detailed understanding of the mechanism of each instability is necessary to interpret the results of numerical simulations and guide the exploration of the parameter space. It is also useful to identify potential numerical artefacts, for example on the competition between SASI and convection \citep{FryerWarren2004,Hanke+2013,Ott+2013}. SASI results from the interaction of pressure and advected perturbations of entropy and vorticity between the shock and the surface of the PNS \citep{Foglizzo+2007,Foglizzo2009,Fernandez+2009,Guilet+2012}. The convective instability in the post-shock region is driven by neutrinos emitted by the cooling PNS: in the region where the absorption of neutrino energy exceeds the losses by neutrino emission, the heating by neutrino absorption creates a negative entropy gradient favourable to convection \citep{Herant+1992}.
%\citep{Bethe1990}. 
The growth of this neutrino-driven convection requires a strong enough neutrino heating such that the buoyancy timescale is shorter than one third of the advection timescale across this region \citep{Foglizzo+2006}. Above a critical heating rate, the fundamental oscillatory mode of SASI becomes the purely growing convective instability that dominates the dynamics of the flow \citep{Yamasaki+2006, Fernandez+2014}.

Most core-collapse simulations and theoretical work neglect the impact of rotation, which is therefore not well known. % \citep{Suwa+2010,Chatzopoulos+2016, fujisawa19, PowellMuller2020}. 
Rotation is however thought to play an important role in at least a small fraction of core-collapse supernovae with more extreme properties, such as superluminous supernovae \citep{woosley10,inserra13} or hypernovae and long gamma-ray bursts \citep{woosley93,metzger11}. It is furthermore possible that rotation plays a less extreme role in a larger fraction of core-collapse supernovae. Theoretical models of stellar evolution constrained by the efficient transport of angular momentum inferred from asteroseismic observations of red giants \citep{Cantiello+2014}, and the observations of pulsar spins \citep{PopovTurolla2012} suggest that the majority of supernova explosions originate from slowly rotating stellar cores, and the rotation frequency $\Omega$ could be as low as $2\times 10^{-3}$~rad/s \citep{MaFuller2019}. 

Rotation rates up to $2$ rad/s are more exceptional but commonly considered to explain extreme events such as hypernovae, superluminous supernovae and GRBs. In this regime, the convective dynamo \citep[e.g.][]{Thompson93,Raynaud+2020} and the magnetorotational instability \citep[e.g.][]{akiyama03,Guilet+2022,reboul-salze22} are expected to amplify efficiently the magnetic field of the PNS. \cite{Raynaud+2020} predicted that magnetar-like magnetic fields can be generated by the strong field branch of the convective dynamo, which takes place at early time for specific angular momentum larger than $4\times10^{15}\, \mathrm{cm^2/s}$. This threshold corresponds to an angular frequency $0.4$~rad/s at $1000$~km in the progenitor. The extraction of rotational energy with such a strong magnetic field can lead to strong magnetorotational explosions \citep[e.g.][]{takiwaki09,Kuroda20,Bugli+2021}. The dynamo timescale being uncertain, the hydrodynamical approximation is often chosen for simplicity to model the majority of supernova.

Beside the generation of magnetic fields, rotation can have several other effects on the shape of the neutrinosphere and on the development of instabilities. The centrifugal force diminishes the action of gravity and results in a larger radius of the neutrinosphere in the equatorial plane. The equatorial decrease in neutrino luminosity and neutrino mean energy is not favourable to the explosion according to axisymmetric simulations \citep{MarekJanka2009}. The lower equatorial temperature favours neutrino heating in the polar region and a bipolar explosion \citep{Suwa+2010}. 

Stellar rotation can be favourable to the development of a vigorous, prograde, spiral SASI mode %dominated by large angular scales $l=1,2$ when neutrino absorption is neglected
\citep{Blondin2007}. This destabilising effect of differential rotation exists even for slow rotation with a negligible centrifugal contribution, as shown by perturbative analyses \citep{YamasakiFoglizzo2008, Walk+2022} and confirmed by numerical simulations \citep{Kazeroni+2017, Blondin+2017}, in which neutrino heating was neglected. However, the driving mechanism of this rotational destabilisation is not understood yet \citep{Walk+2022}.

Classical studies of the effect of rotation on convection considered a rotation axis aligned with gravity. In a viscous fluid with thermal diffusion, the critical Rayleigh number defining the onset of thermal convection is increased by rotation \citep{Chandrasekhar1961,Rossby1969,Wedi2021}. The studies of convection that considered the impact of differential rotation \citep{FeudelFeudel2021} did not involve radial advection, which is crucial in the supernova case.
According to the Solberg-Høiland criterion, the development of axisymmetric convection can be stabilised by rotation if the specific angular momentum increases outward \citep{EndalSofia1978}. In axisymmetric simulations of stellar core-collapse, this effect produces less vigorous convective motions in the equatorial plane, and results in later-time explosions which are weaker at the equator than at the poles \citep{FryerHeger2000}. These 2D results seemed to be confirmed in 3D for the fastest spinning progenitors \citep{FryerWarren2004} ($\Omega \sim 4.1$ rad/s at $1000$ km), in a regime where centrifugal effects can be dominant at reducing both the effective gravity and the neutrino luminosity in the equatorial plane, and thus a reduced buoyancy compared to the polar region. 

Estimating the effect of modest rotation in the equatorial region of post-shock convection is less obvious when centrifugal effects are small, remembering that inward accretion produces a uniform profile of specific angular momentum.
With a rotation rate of $\Omega \sim 1.3$ rad/s at $1000$ km, core-collapse simulations showed an earlier onset of neutrino-driven convection that produced a stronger explosion in the equatorial plane, earlier than the non-rotating case \citep{Nakamura+2014}.

Even a modest amount of rotational kinetic energy $T$ compared to the potential energy $|W|$ can trigger a spiral instability known as “low-$T/|W|$ instability” in the interior of isolated neutron stars \citep{Shibata+2002, Watts+2005, PassamontiAndersson2015}. The mechanism of this instability relies on the extraction of energy and angular momentum from internal regions rotating faster than the spiral pattern towards external region rotating slower \citep{Cairns1979, SaijoYoshida2006}. A similar instability has been observed in 3D simulations of stellar core-collapse, where it enhances the energy transport from the PNS to the shock and can lead to stronger explosions \citep{Ott+2005,CerdaDuran+2007,Takiwaki+2016,Takiwaki+2021}.

The interplay of centrifugal effects, SASI, convection and the low-$T/|W|$ instability can be difficult to disentangle in numerical simulations considering the diversity of progenitors and numerical approximations \citep{Ott+2008}. Our current understanding relies on a very sparse sampling of this diversity. During the collapse of a $27M_{\odot}$ progenitor, the dynamic of the shock is driven by SASI and the convection for small rotation rates, and dominated by the low-$T/|W|$ instability for high enough rotation rates ($\Omega = 2 $ rad/s) \citep{Takiwaki+2021}. The enhancement of SASI by differential rotation can compensate for the loss of neutrino energy due to the centrifugal force during the collapse of a $15M_{\odot}$ progenitor \citep{Summa+2018}. The gravitational wave (GW) analysis of these simulations can help identify the physical processes, such as the enhancement of SASI for high rotation rates \citep{Andresen+2019}. The 3D simulations of stationary accretion by \cite{Iwakami+2014} considered a range of rotation rates, mass accretion rates, and neutrino luminosities. The structure of the dominant instability and the observed patterns were classified according to three main categories: spiral, buoyant bubbles, or spiral and buoyant bubbles. However, the simultaneous variation of both rotation and neutrino luminosity made it difficult to disentangle their respective effects. 

%How we complete previous work
Previous studies of the impact of rotation on SASI \citep{YamasakiFoglizzo2008,Blondin+2017} did not consider the effect of heating. Conversely, perturbative studies on the effect of heating on both SASI and the convective instability did not consider rotation \citep{Yamasaki+2006,Fernandez+2014}.
To have a better understanding of the effect of each parameter on the growth of instabilities, we use a linear analysis, varying the rotation and the heating rate separately. Doing this, we are able to disentangle the effect of these parameters in the linear regime. By focusing on the accretion region above the surface of the PNS, we do not include the interaction with the low-$T/|W|$ and the convective instabilities developing inside the PNS. The aim of this paper is thus to study the impact of rotation on the onset of neutrino-driven convection and its interplay with SASI, when both the neutrino heating and the rotation rate are varied. In Sect.~\ref{sec_methods}, we detail the numerical set-up and define the stationary and perturbed flows. We study in Sect.~\ref{sec_results} the effect of rotation on SASI, convection and their interplay. Finally, we focus in Sect.~\ref{sec_GW} on the features that might be observable in a gravitational wave signal coming from an exploding supernova.

\section{Methods\label{sec_methods}}

\subsection{Numerical setup}

To study the growth of the convective and SASI instabilities, we solve numerically the system of perturbed equations corresponding to an idealised model of stationary accretion of a perfect gas, in spherical geometry restrained to the equatorial plane as in \cite{Walk+2022}, using the coordinates $(r,\phi)$. Our setup is an adaptation of the linear analysis of \cite{Fernandez+2014} to include rotation.
The parameters we vary are the reference shock radius $r_{{\rm sh}0}$ obtained without neutrino heating, dissociation and rotation, the rate $\varepsilon$ of nuclear dissociation across the shock, the neutrino luminosity and the specific angular momentum $J$ reaching the surface of the PNS.

\subsection{Stationary flow}

%stationary flow
In spherical geometry, the system of stationary equations describing the conservation of mass, the entropy profile $S(r)$, the conservation of angular momentum, and the profile of the Bernoulli parameter in the equatorial plane are: 
\begin{eqnarray}
    \diffp{}{r}\left(\rho v r^2\right) &=& 0,\\
    \diffp{S}{r} &=& \dfrac{\mathcal{L}}{Pv},\\
    \diffp{J}{r} &=& 0, \label{eq:ang_mom}\\
    \diffp{}{r}\left(\dfrac{v^2}{2} + \dfrac{J}{2r^2} + \dfrac{c^2}{\gamma-1} - \dfrac{GM}{r}\right) &=& \dfrac{\mathcal{L}}{\rho v},
\end{eqnarray}
where $G$ is the universal gravity constant and $M$ is the mass of the PNS,
$v$ the radial velocity and $c$ the sound speed in the gas with pressure $P$ and density $\rho$. The self-gravity of the infalling matter is neglected compared to that of the PNS.
Non-adiabatic heating and cooling processes are described by a local function $\mathcal{L} \equiv \mathcal{L}_{\rm h} + \mathcal{L}_{\rm c}$, as in \cite{Fernandez+2014}. The cooling function $\mathcal{L}_{\rm c}$ used in \cite{HouckChevalier1992} is a parametric function of $P$ and $\rho$:
\begin{equation}
    \mathcal{L}_{\rm c} = - A_{\rm c} \ \rho^{\beta-\alpha}P^\alpha.
\end{equation}
We use $\alpha = {3/2}$ and $\beta={5/2}$ as in \cite{BlondinMezzacappa2006}, \cite{Foglizzo+2007}, \cite{YamasakiFoglizzo2008}, \cite{Fernandez+2009}, \cite{Fernandez+2014}, \cite{Guilet+2012} and \cite{Blondin+2017}. %The normalisation constant $A_{\rm c}$ describes the micro-physical processes responsible for the cooling by neutrino emission. %Its value is determined so that the radial velocity vanishes at the surface of the PNS. 
Using a dimensional analysis, we express $A_{\rm c}$ as a function of the PNS radius $r_{\rm PNS}$, the surface gravity $GM/r_{\rm PNS}^2$ and the mass accretion rate $\dot{M}$:
\begin{eqnarray}
A_{\rm c} = \Tilde{A}_{\rm c} \times \dot{M}^{1-\beta} \left({GM\over r_{\rm PNS}^2}\right)^{1-\alpha +{\beta\over2}} r_{\rm PNS}^{{5\beta\over2}-2-\alpha},
\end{eqnarray}
where $\Tilde{A}_{\rm c}$ is a dimensionless quantity. The value of $\Tilde{A}_{\rm c}$ sets the value of the shock radius $r_{\rm sh}$ without dissociation, rotation, or heating. $\Tilde{A}_{\rm c}$ does not vary when these parameters are changed. 
The heating function $\mathcal{L}_{\rm h}$ is expressed as
\begin{equation}
    \mathcal{L}_{\rm h} = A_{\rm h} \dfrac{\rho}{r^2}.
\end{equation}
The normalisation constant $A_{\rm h}$ is proportional to the neutrino luminosities and the neutrino opacities per unit mass. It is varied as a free parameter in our model. The heating and cooling functions are effective in the post-shock region and are turned off above the shock, as in \cite{Fernandez+2014}.

%---------------------
%rotation
In order to study the influence of stellar rotation on the growth of the instabilities, we vary parametrically the specific angular momentum $J$. Its dimensionless measure $j$ is defined using the Keplerian specific angular momentum at the PNS surface
\begin{equation}
    j \equiv \frac{J}{\left(GMr_{\rm PNS}\right)^{1/2}}.
\end{equation} 
We note that $j^2$ measures the ratio of the centrifugal and gravitational forces at the PNS radius:
\begin{equation}
    \frac{J^2}{r_{\rm PNS}^3}\frac{r_{\rm PNS}^2}{GM} = j^2.
    \label{eq:forces}
\end{equation}
In our analysis, we vary $j$ from $0$ to $0.5$ which corresponds to a maximum centrifugal force equal to $25\%$ of the gravitational force at the PNS boundary. The angular momentum at $50\%$ of the Keplerian rotation can be expressed as
\begin{equation}
    J = 1.5\times 10^{16}{~\rm cm^2/s} \ \frac{j}{0.5} \left(\frac{r_{\rm PNS}}{50 ~{\rm km}} \frac{M}{1.4~{\rm M_\odot}}\right)^{1/2}
\end{equation}
which corresponds to an angular frequency $\Omega = 1.5 {~\rm rad/s}$ at a reference radius $r= 1000 {~\rm km}$ in the progenitor. For comparison, the strong dynamo branch described in \cite{Raynaud+2020} could be generated for $j\gtrsim 0.13$. Our hydrodynamical study neglecting magnetic fields assumes that dynamo processes are too slow to interfere.

Angular momentum conservation leads to a uniform specific angular momentum in the stationary flow (Eq.~\ref{eq:ang_mom}). The profile of angular frequency is therefore
\begin{equation}
\Omega(r) = \frac{J}{r^2} = 150 ~{\rm rad/s} \ \frac{j}{0.5}\ \left(\frac{100~{\rm km}}{r}\right)^2.
\end{equation}
%---------------------

\begin{figure}[h]
    \centering
    \includegraphics[width=\hsize]{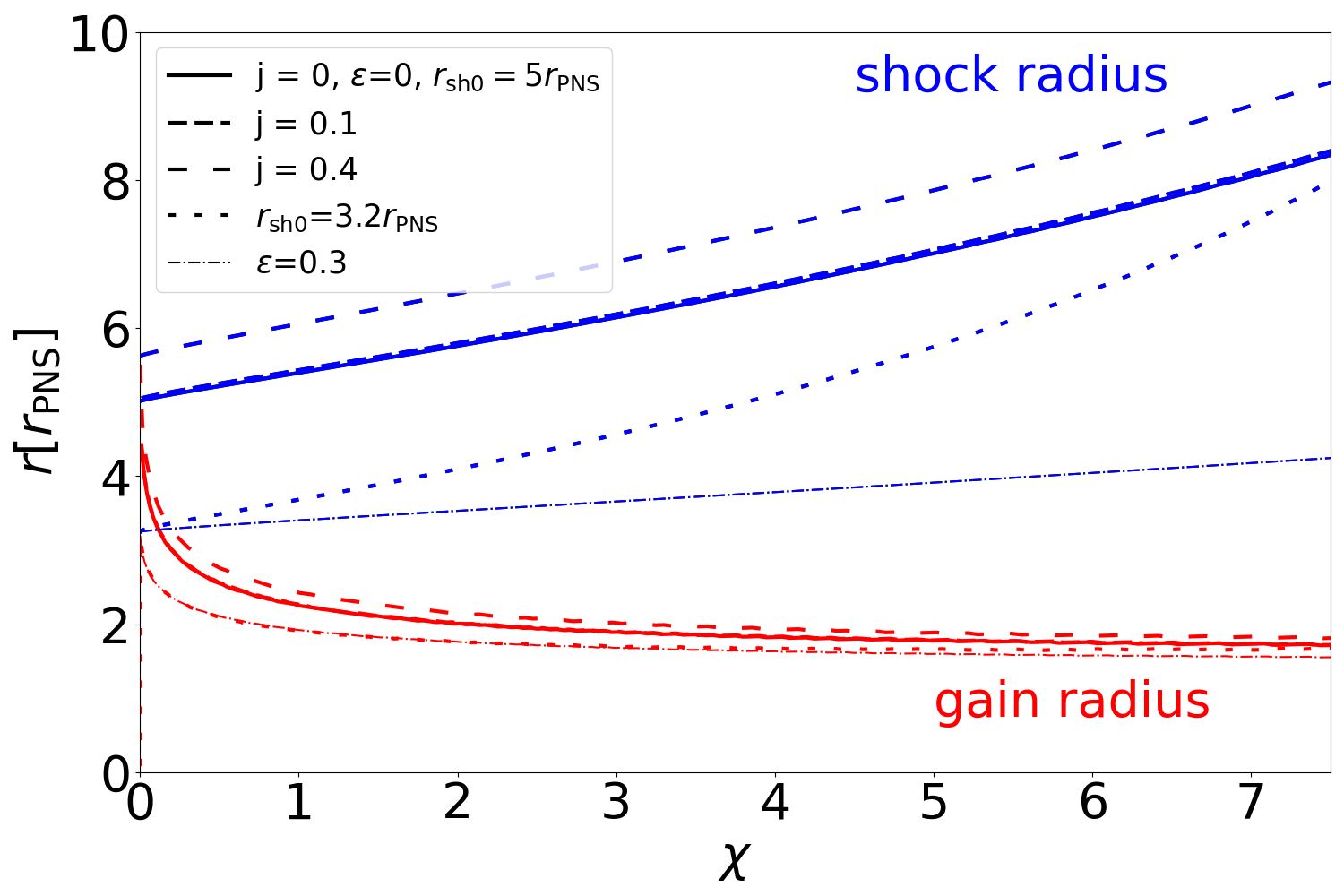}
    \caption{Evolution of the shock (blue) and gain (red) radii as a function of the convection parameter $\chi$. Unless stated otherwise, the values of the parameters are $r_{{\rm sh}0}=5r_{\rm PNS}$, $\varepsilon=0$ and $j=0$. The case $(j,\varepsilon)=(0,0)$ and $r_{{\rm sh}0}=3.2r_{\rm PNS}$ (dashed lines) is chosen such that the shock radius without heating coincides with the case with $(j,\varepsilon)=(0,0.3)$ and $r_{{\rm sh}0}=5 r_{\rm PNS}$. As expected, the shock radius increases with $\chi$ and the introduction of rotation increases the shock radius when the cooling normalisation $A_{\rm c}$ is conserved. Note that for the moderate value of angular momentum $j=0.1$, the stationary flow is almost unaffected by rotation.}
    \label{rsh_rg(chi)}
\end{figure}

%parameters 
In our model, the dissociation parameter $\varepsilon$ corresponds to the fraction of specific kinetic energy of the incoming matter that is used to photo-dissociate the iron nuclei. $\varepsilon$ is expressed in units of the specific kinetic energy $v_{{\rm ff}0}^2/2$ associated to free-fall at the reference radius $r_{{\rm sh}0}$. The Mach number is imposed so that $\mathcal{M}_1 = 5$ above the shock, without heating, rotation, or dissociation. The Bernoulli parameter is set to zero above the shock, and the adiabatic index is set to $\gamma=4/3$. The default values of the shock radius and the dissociation rate are $r_{{\rm sh}0} = 5 r_{\rm PNS}$ and $\varepsilon = 0$. %The index $0$ stands for the shock radius when $(A_{\rm h}, \varepsilon, J)$ = ($0,0,0$). 
For each figure, the value $r_{{\rm sh}0} = 5 r_{\rm PNS}$ is used and the only exception is the use of $r_{{\rm sh}0} = 3.2 r_{\rm PNS}$ for comparison in Figs.~\ref{rsh_rg(chi)} and \ref{disso_cool}. This specific value is chosen such that the shock radius without heating coincides with the case with $(j,\varepsilon)=(0,0.3)$ and $r_{{\rm sh}0}=5 r_{\rm PNS}$. The value of $r_{\rm sh}$ specified in each figure corresponds to the shock radius taking into account the neutrino heating, dissociation and rotation associated to the parameters $(A_{\rm h}, \varepsilon, j)$. The shock radius decreases when dissociation increases. As the rotation or the heating rate $A_{\rm h}$ is increased, the shock radius increases (blue curve in Fig.~\ref{rsh_rg(chi)}) and the corresponding Mach number immediately above the shock decreases. 

Distances are normalised with either the PNS radius $r_{{\rm PNS}}$ or the reference shock radius $r_{{\rm sh}0}$. Densities are normalised with the density immediately above the shock $\rho_{10}$, and velocities with the free-fall velocity $v_{{\rm ff}0}$ at $r_{{\rm sh}0}$. These units are chosen for $(A_{\rm h}, \varepsilon, j)=(0,0,0)$ and they do not vary when $(A_{\rm h}, \varepsilon, j)$ are varied. With these units, the heating normalisation constant is written
\begin{equation}
    A_{\rm h} = \tilde{A_{\rm h}} v_{{\rm ff}0}^3 r_{{\rm sh}0},
\end{equation}
where $\tilde{A_{\rm h}}$ is a dimensionless parameter characterising the heating rate.
Another dimensionless measure of the heating rate is the convection parameter $\chi$ defined in \cite{Foglizzo+2006} as 
\begin{equation}
    \chi \equiv \int_{r_{\rm g}}^{r_{\rm sh}} N(r)\dfrac{dr}{|v|} ,
\end{equation}
where $r_{\rm g}$ is the gain radius above which absorption of neutrino energy exceeds the losses by neutrino emission, and $N(r)$ is the local Brunt-Väisälä frequency:
\begin{equation}
    N \equiv \left(\dfrac{\gamma-1}{\gamma}g\nabla S\right)^{1\over2}.
\label{definition_N}
\end{equation}
In this equation, the gravity term $g$ was not corrected by the centrifugal force for the sake of simplicity, which is acceptable because of the low value of the ratio of those two forces in the parameter domain that we explore. Indeed, Eq.~\eqref{eq:forces} shows that even for the fastest rotation rate considered ($j=0.5$), the ratio of those forces reaches only $\sim7\%$ in the middle of the post-shock region.

We thus use $(r_{{\rm sh}0},\chi,\varepsilon,j)$ to explore the 4D parameter space associated to the mass accretion rate, neutrino luminosity, dissociation rate and rotation.

Figure~\ref{rsh_rg(chi)} illustrates the dependence on neutrino heating of the shock and gain radii for typical rotation and dissociation parameters. %indicates that the gain radius varies little with the strength of neutrino heating or the dissociation rate when the heating rate is high enough. 
The gain radius is located near the shock for small values of $\chi$, approaches twice the PNS radius for $\chi\sim 2$ and remains constant for $\chi\gg 2$ while the radius of the stationary shock is pushed further out by the strong heating of the matter. 
%For $\chi \gtrsim 2$, the gain radius stalls at $\sim 2r_{\rm PNS}$ and the size of the gain region increases proportionally to the shock radius.

The stronger the dissociation, the lower the energy after the shock. This leads to lower velocities of sound and matter than in the case without dissociation. As a result, the integrated intensity of the cooling function needed to decelerate across the shocked region is reduced. As $\Tilde{A}_{\rm c}$ is determined with $\varepsilon=0$, the size of the shocked region decreases when the dissociation rate increases, as observed in Fig.~\ref{rsh_rg(chi)}.

The centrifugal force increases the shock radius by $\simeq 15\%$ for the fast rotation rate $j=0.4$. On the other hand, the impact of rotation is barely visible in this figure for a moderate rotation rate $j=0.1$ because of the quadratic dependence of the centrifugal force with the rotation frequency as measured by Eq.~\eqref{eq:forces}. With $j=0.1$, we obtain a centrifugal effect of only $j^2=1\%$. However, we will see in the next section that this modest rotation is large enough to have significant effects on SASI and the convective instability.

\subsection{Perturbed flow}
%---------------------
%perturbations
For the perturbed system, we use the same set of perturbed variables as in \cite{YamasakiFoglizzo2008}. $\delta f$, $\delta h$, $\delta S$, and $\delta q$ are defined as follows:
\begin{eqnarray}
    \delta f &\equiv& v \delta v + \dfrac{J}{r}\delta v_\phi + \dfrac{2c}{\gamma -1}\delta c -\delta q , \\
    \delta h &\equiv& \dfrac{\delta v}{v} + \dfrac{\delta \rho}{\rho} , \\
    \delta S &\equiv& \dfrac{2}{\gamma-1}\dfrac{\delta c}{c} - \dfrac{\delta \rho}{\rho} , \\
    \delta q &\equiv& \delta\left(\int \dfrac{\mathcal{L}}{\rho v} dr\right).
\end{eqnarray}
The time and angular dependence of the perturbations are assumed to follow the form ${\rm exp}\left(-i\omega t + im\phi\right)$. 
Their radial dependence is then governed by the same differential equations as Eqs.~(6-11) in \cite{YamasakiFoglizzo2008} with $k_z=0$, despite the different geometry (spherical-equatorial instead of cylindrical) and the inclusion of neutrino heating in the local function ${\cal L}$:
\begin{eqnarray}
    {{\rm d}\over{\rm d} r}{\delta f\over\omega} &= &\dfrac{i c^2}{v \left(1-\mathcal{M}^2\right)} \left\{ \mathcal{M}^2 \left( \delta h -\dfrac{\omega '}{c^2}\dfrac{\delta f }{\omega}\right)\right.\\
    &&\left.+ \left[1+\left(\gamma-1\right)\mathcal{M}^2\right]\dfrac{\delta S}{\gamma} - \dfrac{\delta q }{c^2}\right\}   ,\nonumber\\
    \diff{\delta h}{r} &= &\dfrac{i \omega}{v \left(1-\mathcal{M}^2\right)}\left( \mu^2\dfrac{\omega '}{c^2}\dfrac{\delta f }{\omega} - \mathcal{M}^2\delta h - \delta S +\dfrac{\delta q}{c^2}\right)   ,\\
    \diff{\delta S}{r} &=  &\dfrac{i\omega'}{v}\delta S + \delta\left(\dfrac{\mathcal{L}}{P v}\right)  ,\\
    \diff{\delta q}{r} &= &\dfrac{i\omega'}{v}\delta q + \delta \left(\dfrac{\mathcal{L}}{\rho v}\right)  ,
\end{eqnarray}
where 
\begin{eqnarray}
    \omega' &\equiv& \omega -\dfrac{mJ}{r^2},\label{doppler_omega}\\
    \mu^2 &\equiv& 1 - {m^2\over r^2}\dfrac{c^2}{\omega'^2}\left(1-\mathcal{M}^2\right).
\end{eqnarray}
In these equations $\omega'$ corresponds to the Doppler shifted value of $\omega$ and is thus a function of radius. Note that this effect of rotation is linear with respect to $\Omega$, whereas the centrifugal force is proportional to $\Omega^2$ as discussed above. A linear dependence allows for strong effects of slow rotation, as will be discussed later.

The boundary conditions at the shock are expressed using conservation laws across a perturbed shock:
\begin{eqnarray}
    {\delta f_{\rm sh}\over\omega} &=&  iv_1 \Delta\zeta \left(1-\dfrac{v_{\rm sh}}{v_1}\right),\\
    \delta h_{\rm sh} &=&  -i\dfrac{\omega'}{v_{\rm sh}} \Delta\zeta \left(1-\dfrac{v_{\rm sh}}{v_1}\right),\\
    \delta S_{\rm sh} &=&  i\dfrac{\omega'v_1}{c_{\rm sh}^2}\Delta\zeta \left(1-\dfrac{v_{\rm sh}}{v_1}\right)^2 - \dfrac{\mathcal{L}_{\rm sh}-\mathcal{L}_1}{\rho_{\rm sh}v_{\rm sh}}\dfrac{\Delta\zeta}{c^2_{\rm sh}} \nonumber\\
  &  & + \left(1-\dfrac{v_{\rm sh}}{v_1}\right)\dfrac{\Delta\zeta}{c^2_{\rm sh}}\left(\dfrac{2v_1v_{\rm sh}}{r_{\rm sh}} + \dfrac{J^2}{r_{\rm sh}^3} + \dfrac{GM}{r_{\rm sh}^2}\right),\label{deltaSsh} \\
    \delta q_{\rm sh} &=& -\dfrac{\mathcal{L}_{\rm sh}-\mathcal{L}_1}{\rho_{\rm sh}v_{\rm sh}} \Delta\zeta ,
\end{eqnarray}
where $\Delta\zeta$ is the shock displacement. The indices “$1$” and “sh” refer to the values above and below the shock, respectively. The only difference with the boundary conditions (12-15) in \cite{YamasakiFoglizzo2008}, who assumed a cylindrical symmetry, is the factor $2$ implied by our choice of spherical-equatorial geometry, in the term $2v_1v_{\rm sh}/r_{\rm sh}$ in Eq.~\eqref{deltaSsh}.

We use the same lower boundary condition as in \cite{YamasakiFoglizzo2008} at the PNS surface, where the radial velocity vanishes.
We use the Newton-Raphson method to find the eigenvalues of this system when $\varepsilon$, $\mathcal{M}_1$, $\chi$, and $r_{\rm sh}$ (or $\tilde{A}_{\rm c}$) are given. This method determines the perturbation growth rate $\omega_i$ and frequency $\omega_r$ associated to the complex eigenfrequency $\omega\equiv\omega_r+i\omega_i$, for each azimuthal number $m$.

In order to identify the effect of the four parameters 
$(r_{{\rm sh}0},\varepsilon,\chi, j)$ on the growth of instabilities, we select a mode and follow its evolution when the parameters are varied. For a given value of $m$, we studied the evolution of each harmonic without rotation and selected the overtone with the highest growth rate, which happens to be the fundamental mode as obtained by \cite{Yamasaki+2006}. We therefore follow the properties of this fundamental mode when the parameters $(r_{{\rm sh}0},\varepsilon,\chi, j)$ are varied. When rotation exceeds $j\sim0.15$, we noticed that some higher harmonics can be slightly more unstable than the fundamental mode, but this effect is marginal.

\section{Results\label{sec_results}}

\subsection{SASI and the convective instability without rotation}

\begin{figure}[h]
    \centering
    \includegraphics[width=\hsize]{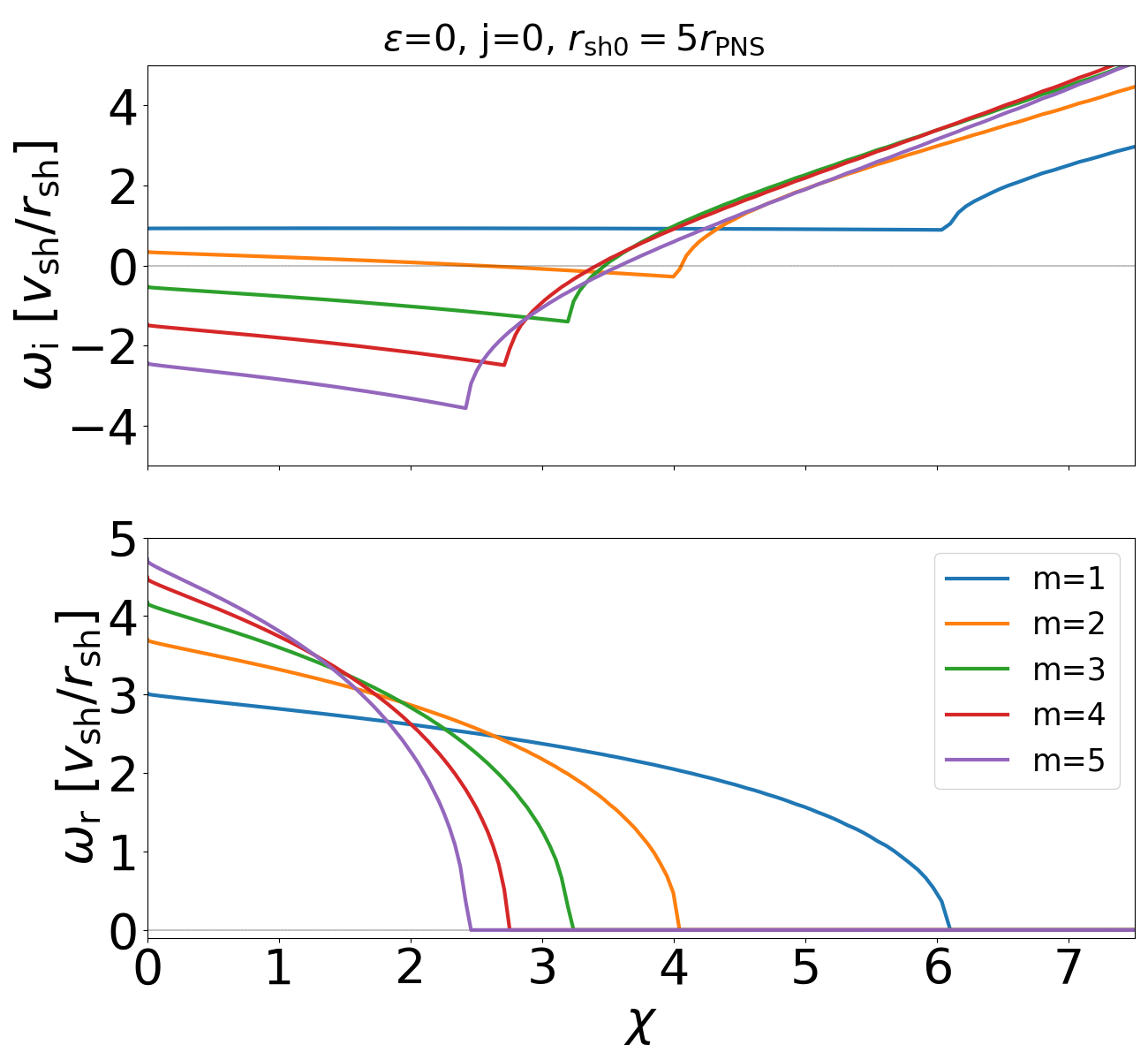}
    \caption{Evolution of the growth rate (upper panel) and frequency (lower panel) of the fundamental modes as a function of the convection parameter $\chi$. Two instability domains can be identified: SASI ($\omega_r\neq$0) for modest neutrino heating and the convective instability ($\omega_r=$0) for stronger heating.}
    \label{wi(chi)}
\end{figure}

Figure~\ref{wi(chi)} shows the effect of the heating parameter $\chi$ on the growth rate and the frequency of the fundamental mode corresponding to several azimuthal numbers $m=1$ to $5$. Similarly to \cite{Yamasaki+2006} and \cite{Fernandez+2014}, for each value of $m$, the instability is oscillatory (like SASI) for modest heating and purely growing (like the convective instability) for stronger heating. The most unstable perturbations in the oscillatory regime have a large angular scale corresponding to a small azimuthal number $m=1$, as expected for SASI. For each mode, the SASI frequency decreases abruptly to zero when the heating rate is increased. The transition to a mode with zero frequency corresponds to a steep increase of the growth rate. The azimuthal scale of the dominant convective modes is smaller (i.e. larger $m$, here $m=4$) than SASI modes. The $\chi$ parameter corresponding to the transition between SASI and convection is $\sim$7\% smaller than expected from \cite{Fernandez+2014}. This difference may be due to their use of an entropy cut-off in the heating/cooling function, ${\rm exp}-(S/S_{\rm min})^2$, 
where $S_{\rm min}$ is the entropy at the PNS surface. Except for this small quantitative difference, the general behaviour is the same as in \cite{Fernandez+2014}.

\begin{figure}[h]
    \centering
    \includegraphics[width=\hsize]{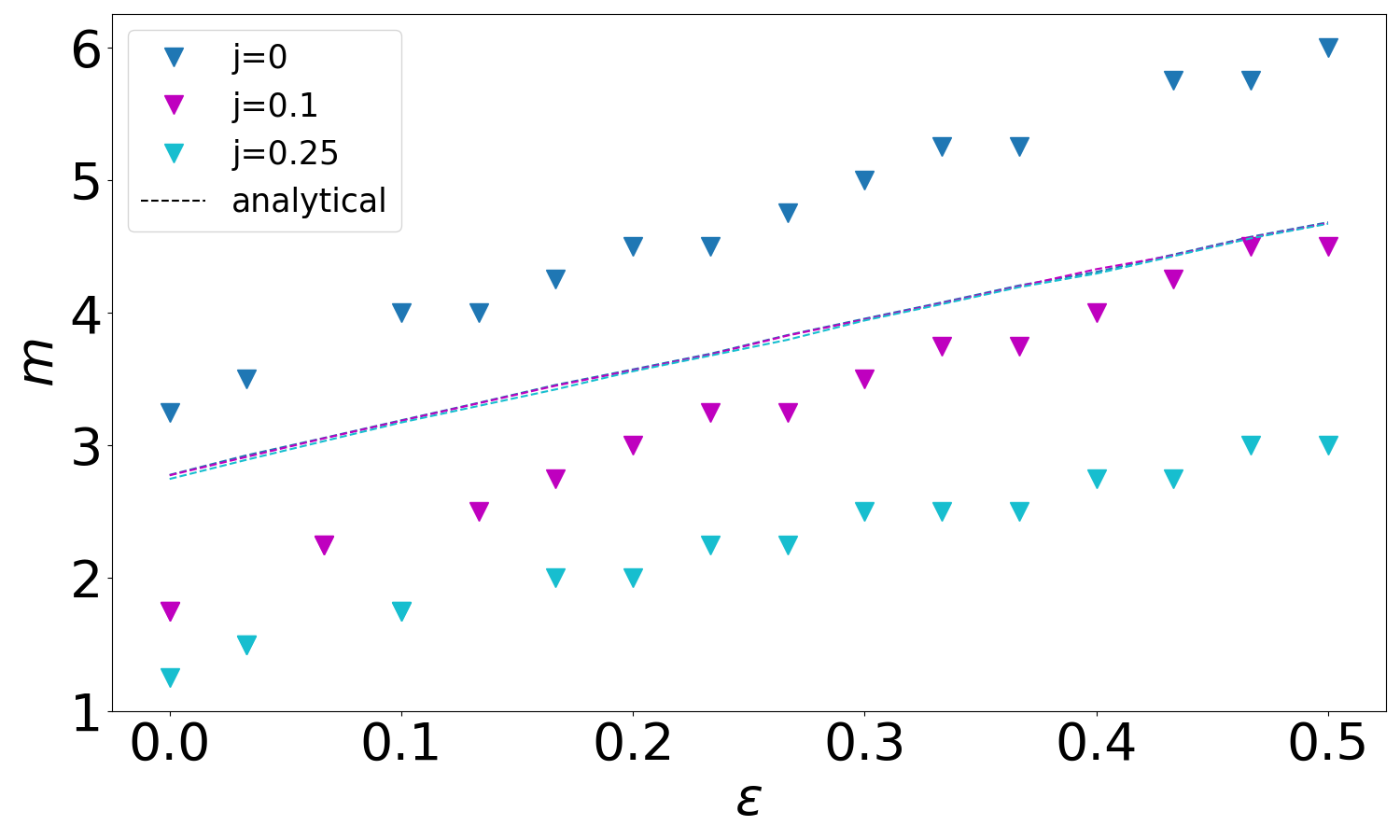}
    \caption{Influence of the dissociation of nuclei across the shock on the azimuthal number of the most unstable fundamental mode $m_{\rm num}$ (triangles) at $\chi=4$ for three different values of the rotation rate. The numerical values of the most unstable modes (triangle) are compared to the analytical predictions of Eq.~\eqref{eq_m} (dashed lines) as a function of the dissociation rate. We allow for quarter-integer values of $m_{\rm num}$ for an easier comparison with Eq.~\eqref{eq_m}.}
    \label{m_dom(eps)}
\end{figure}

To illustrate the influence of the dissociation parameter $\varepsilon$ %and the specific angular momentum $J$ 
on the growth of the convective instability, we fixed the value $\chi=4$ such that the dominant mode is convective for every dissociation rate. Fig.~\ref{m_dom(eps)} compares the azimuthal number $m_{\rm num}$ corresponding to the most unstable mode obtained numerically (triangles), to the expected value estimated analytically by \cite{Foglizzo+2006} (dotted line):
\begin{equation}
    m_{\rm ana} = \dfrac{\pi}{2} \dfrac{r_{\rm sh}+r_{\rm g}}{r_{\rm sh}-r_{\rm g}}.
\label{eq_m}
\end{equation}
The analytical formula is in approximate agreement with $m_{\rm num}$ and reproduces reasonably well its increase with the dissociation parameter $\varepsilon$. This dependence on $\varepsilon$ can therefore be understood using Eq.~\eqref{eq_m}, which is a geometric formula based on the assumption that convective cells have a circular shape, radially centred in the gain region. When dissociation is increased, the shock radius decreases while the gain radius is barely affected (see Fig.~\ref{rsh_rg(chi)}). The smaller radial size of the convective zone therefore leads to smaller-scale convective cells. Beyond the approximate agreement, Fig.~\ref{m_dom(eps)} shows that the analytical formula tends to underestimate the value of $m_{\rm num}$ by $15-25\%$, which can be interpreted as follows. 
Eq.~\eqref{eq_m} neglects the spherical geometry of the star and the fact that the local value of the Brunt-Väisälä frequency is highest near the gain radius such that the width of its radial profile is only a fraction of the gain region size. As a result, the limiting scale for the convective cells is smaller than the whole gain region and comparable to the size of the most buoyant region, as can be seen in Fig.~\ref{structure}.
This can explain the offset between the blue triangles and the blue curve in Fig.~\ref{m_dom(eps)}, indicating that $m_{\rm ana}$ over-estimates the angular size of the dominant convective scale. 

We also varied $r_{{\rm sh}0}$ between $2$ and $4.5$ times the PNS radius by changing the cooling intensity $\Tilde{A}_{\rm c}$ and adapting the pre-shock Mach number $\mathcal{M}_1$ such that $\mathcal{M}_1=5$ for $r_{\rm sh}=5r_{\rm PNS}$, $\varepsilon = 0$. For this variable, the agreement between $m_{\rm num}$ and $m_{\rm ana}$ is similar to the one obtained in Fig.~\ref{m_dom(eps)} without dissociation.

\subsection{Influence of stellar rotation}

In this section, we characterise the dependence of the growth rate and the oscillation frequency on the rotation rate for both SASI and the convective instability. We highlight the existence of three rotation regimes, reflected by the properties of their eigenfrequencies and structures. For small rotation rates ($j\lesssim0.1$, Sects.~\ref{sec:sasi_rotation} and \ref{sec:convection_rotation}), we identify the development of SASI or the convection, depending on the convection parameter. For intermediate rotation rates ($j \in [0.1,0.3]$, Sect.~\ref{sec:mix_modes}), the distinction between the two instabilities is no longer possible and we observe the growth of mixed modes. For high rotation rates ($j \gtrsim 0.3$, Sect.~\ref{part:high_rot}), the effect of the $\chi$ parameter on both the frequency and the growth rate becomes negligible and a rotational instability arises.

\subsubsection{Effect of rotation and heating on SASI}
    \label{sec:sasi_rotation}
The growth rate of SASI is expected to increase approximately linearly with the rotation rate, and the angular scale of the dominant mode is expected to diminish when the rotation rate increases. The corotation radius, defined as
\begin{equation}
    r_{\rm corot} \equiv \left(\dfrac{mJ}{\omega_r}\right)^{1/2},
\label{definition_rcorot}
\end{equation} 
is also expected to increase with the rotation rate. 
These results were obtained without neutrino heating using a perturbative analysis and numerical simulations both in cylindrical geometry \citep{YamasakiFoglizzo2008, Kazeroni+2017} and in spherical geometry \citep{Blondin+2017, Walk+2022}. 

\begin{figure}[h]
    \centering
    \includegraphics[width=\columnwidth]{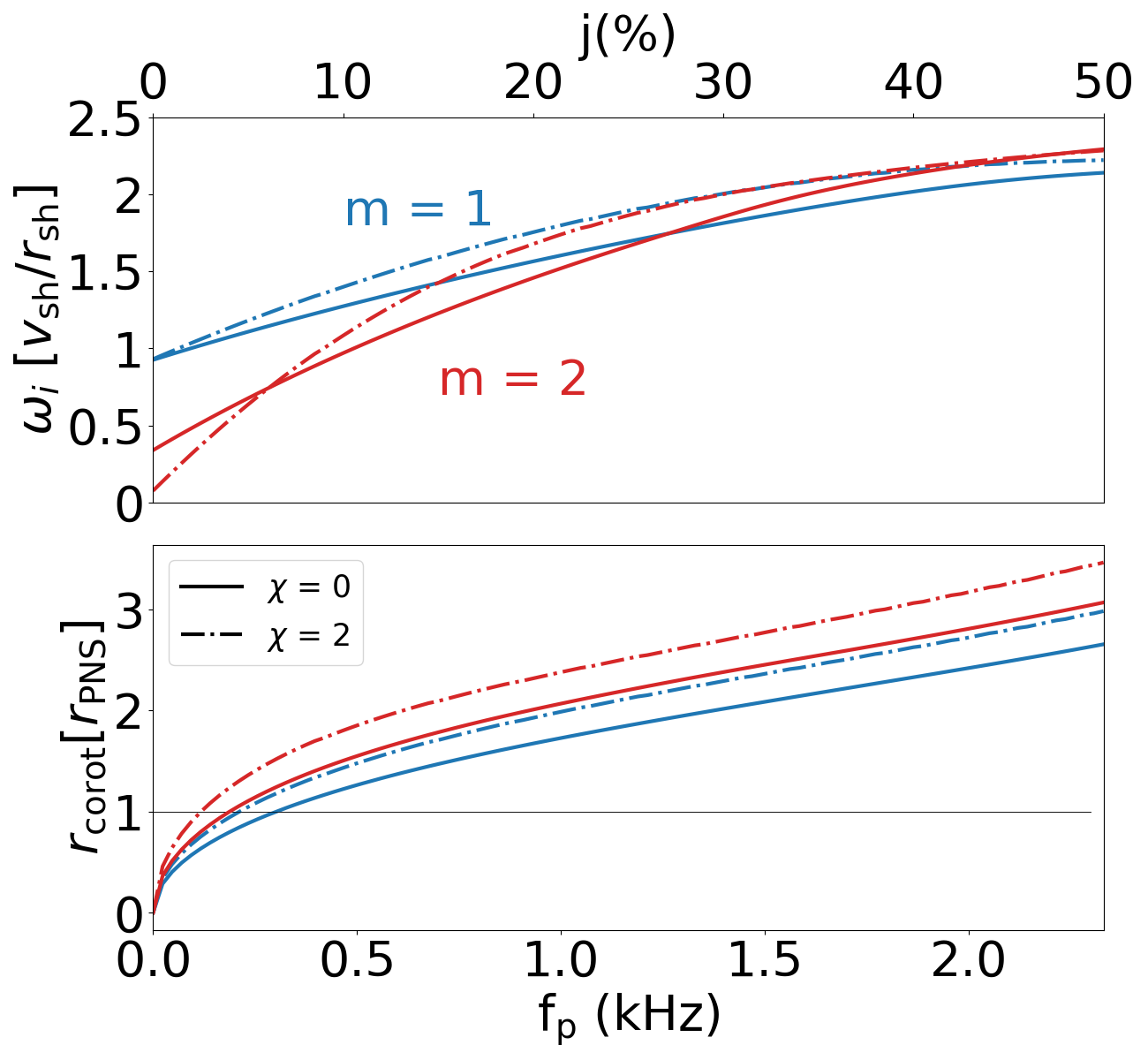}%r_corot(fp)unit_vshrsh.png}
    \caption{Evolution of the growth rate $\omega_i$ in units of $v_{\rm sh}/r_{\rm sh}$ (top), and the corotation radius in units of $r_{\rm PNS}$ (bottom) as a function of the rotation rate, measured as in \cite{YamasakiFoglizzo2008} with $f_{\rm p}$, the pulsar frequency at $10$ km. The azimuthal numbers $m=1$ (blue) and $m=2$ (red) are displayed for $\chi=0$ (solid line), and $\chi=2$ (dash dotted line). To allow for a comparison with Fig.~1 of \cite{YamasakiFoglizzo2008}, we considered the same parameter $r_{\rm sh}=5r_{\rm PNS}$ with a strong adiabatic shock.}
    \label{rcorot(fp)}
\end{figure}

Our perturbative analysis in spherical-equatorial geometry, shown in Fig.~\ref{rcorot(fp)}, confirms these results for $\chi=0$ and measures the impact of neutrino heating for $\chi=2$.
The lower plot in Fig.~\ref{rcorot(fp)} shows the same dependence of the corotation radius on the rotation rate as in \cite{YamasakiFoglizzo2008} despite the different geometry. We note that increasing the rotation rate introduces a corotation radius $r_{\rm corot}>r_{\rm PNS}$ associated to the mode $m=2$ for a lower rotation rate than the mode $m=1$. This hierarchy is conserved when neutrino heating is taken into account. In addition, in \cite{YamasakiFoglizzo2008}, the frequencies for which $r_{\rm corot} = r_{\rm PNS}$ are $f_{p2} \sim 0.15$~kHz and $f_{p1} \sim 0.25$~kHz for $m=2$ and $m=1$, respectively. These values are very close to our spherical-equatorial results, where $f_{p2} \sim 0.17$~kHz and $f_{p1} \sim 0.28$~kHz.

%Comparaison Y08, effect of 'cold' rotation

As in \cite{YamasakiFoglizzo2008}, the growth rate of the mode $m=1$, dominant without rotation, becomes inferior to the growth rate of the mode $m=2$ for sufficiently fast rotation rates. We find that this behaviour is robust when neutrino heating is taken into account, as illustrated in Fig.~\ref{rcorot(fp)} for $\chi=2$. 

\begin{figure}[h]
\centering
    \includegraphics[width=\columnwidth]{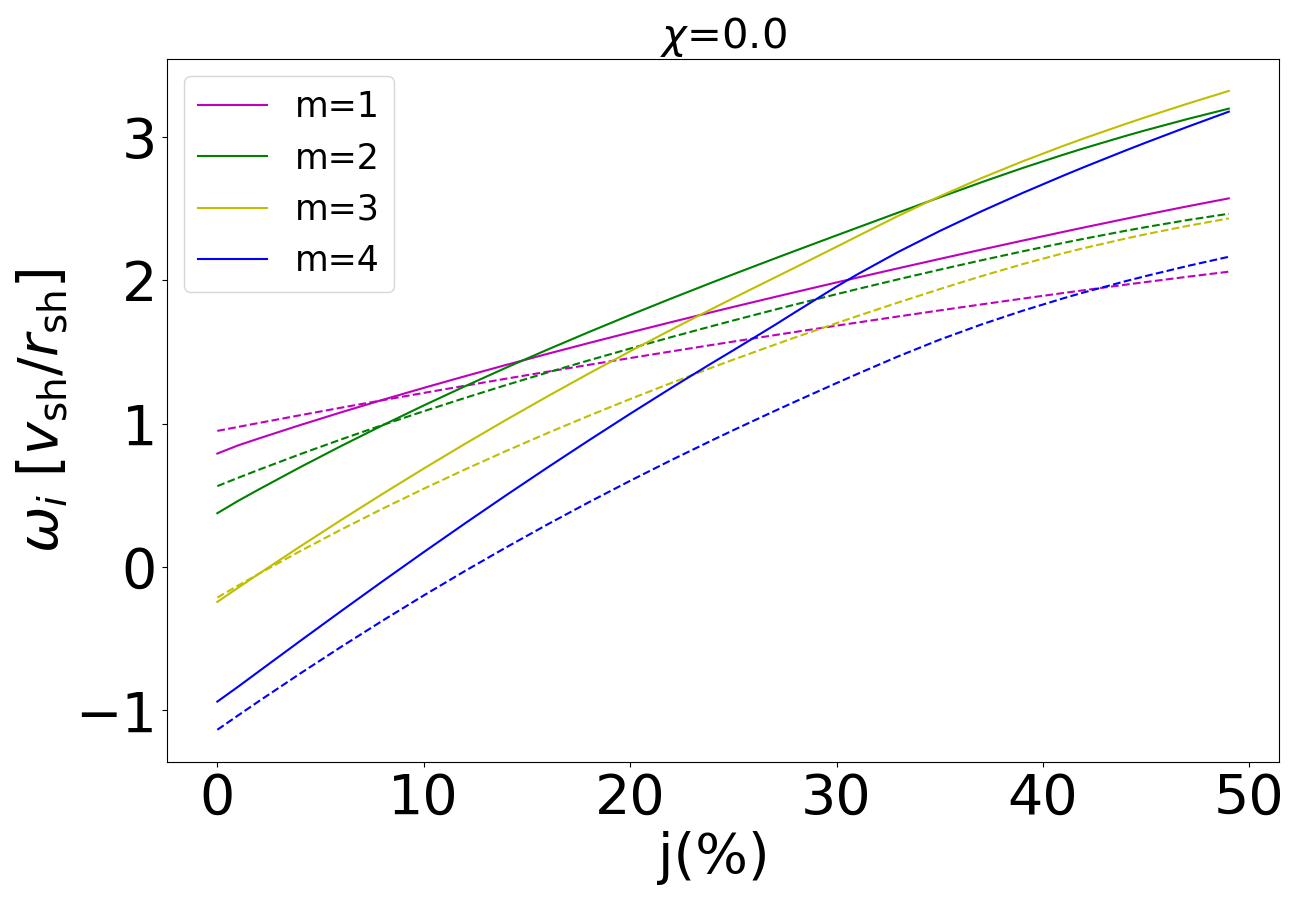}
    \includegraphics[width=\columnwidth,trim={0 0 0 1.5cm},clip]{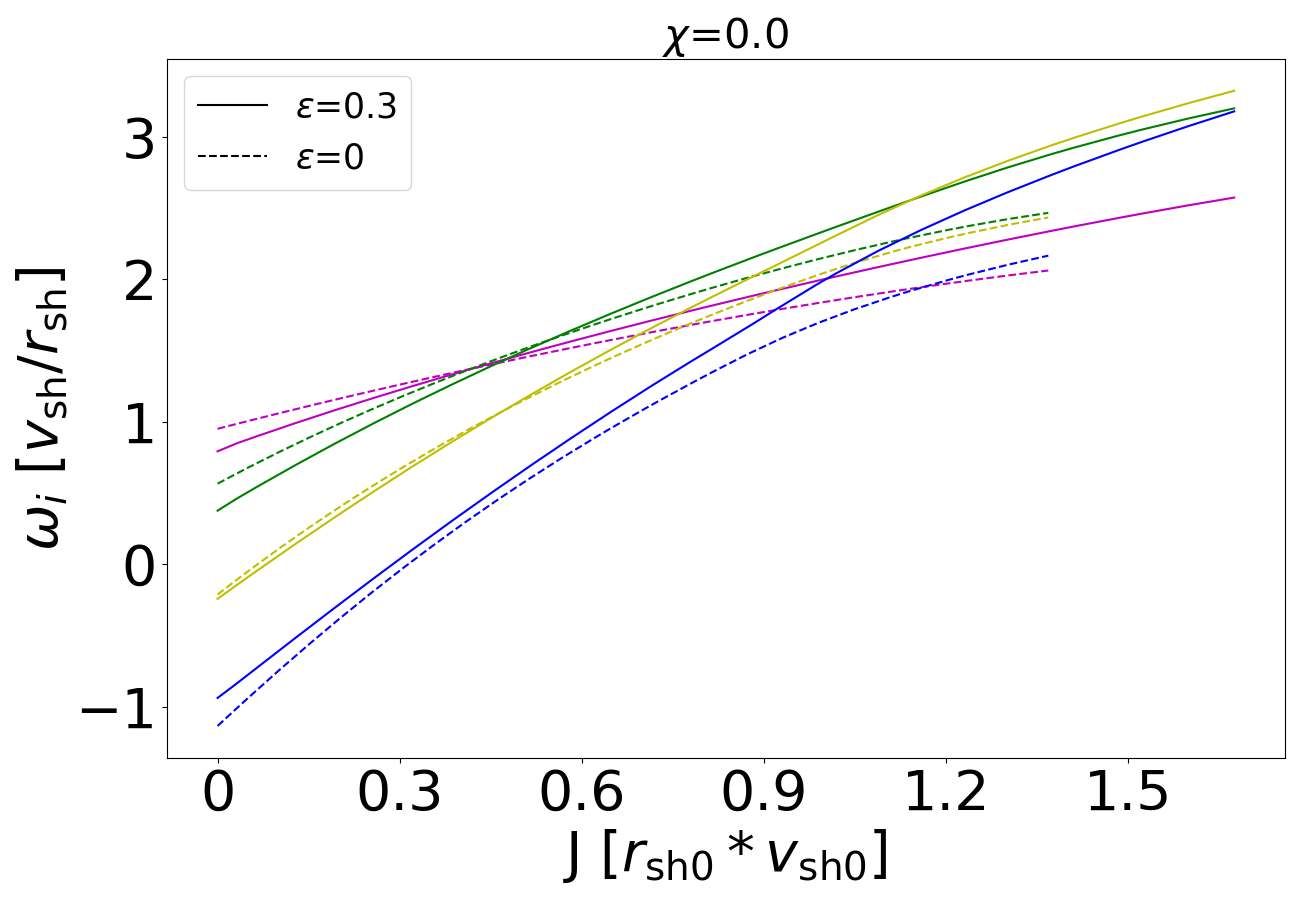}
    \caption{Influence of rotation on the growth rate of SASI with dissociation (solid lines) and without dissociation (dashed lines). The bottom panel presents the rotation rate normalised with a proxy for the advection timescale from the shock to the PNS calculated for $(\chi,j)=(0,0)$. Other parameters are $\chi=0$ and $r_{\rm sh}\left(j=0\right)=3.2r_{\rm PNS}$. }
    \label{disso_cool}
\end{figure}

In the absence of dissociation and heating (solid lines in Fig.~\ref{rcorot(fp)}), the observed transition with the mode $m=2$ dominating the instability for $j\in[0.3,0.5]$ is consistent with the results of \cite{Blondin+2017} where the mode $m=2$ dominates for $j\sim0.36$ (corresponding to $h=0.115$ in their units). % --> h = j*sqrt(rpns/(2rsh))
When neutrino heating is taken into account, Fig.~\ref{rcorot(fp)} shows that the transition from $m=1$ to $m=2$ requires a slightly larger amount of rotation. On the other hand, this transition requires a smaller amount of rotation when the shock radius is smaller or when dissociation is taken into account (Fig.~\ref{disso_cool}). The influence of dissociation can be explained by its effect on the advection timescale. For a given shock radius (kept constant in Fig.~\ref{disso_cool}), the matter velocity decreases with $\varepsilon$ because of the larger compression through the shock. As a consequence, the advection timescale $r_{\rm sh}/v_{\rm sh}$ increases with $\varepsilon$ and the frequency of the SASI modes decreases. A slower rotation is therefore sufficient to be comparable to the SASI frequency, which can explain the steeper effect of rotation. This interpretation is confirmed by the bottom panel of Fig.~\ref{disso_cool} where the rotation rate has been normalised using the advection timescale. This renormalisation indeed collapsed the curves with and without dissociation, at least for the small rotation rates. Note, however, that this argument cannot explain the dependence on the shock radius of the rotational destabilisation of SASI, which remains mysterious.

\subsubsection{Effect of rotation on the convective instability}
\label{sec:convection_rotation}
%-----effect of rotation+heating

\begin{figure}[h]
    \centering
    \includegraphics[width=\hsize]{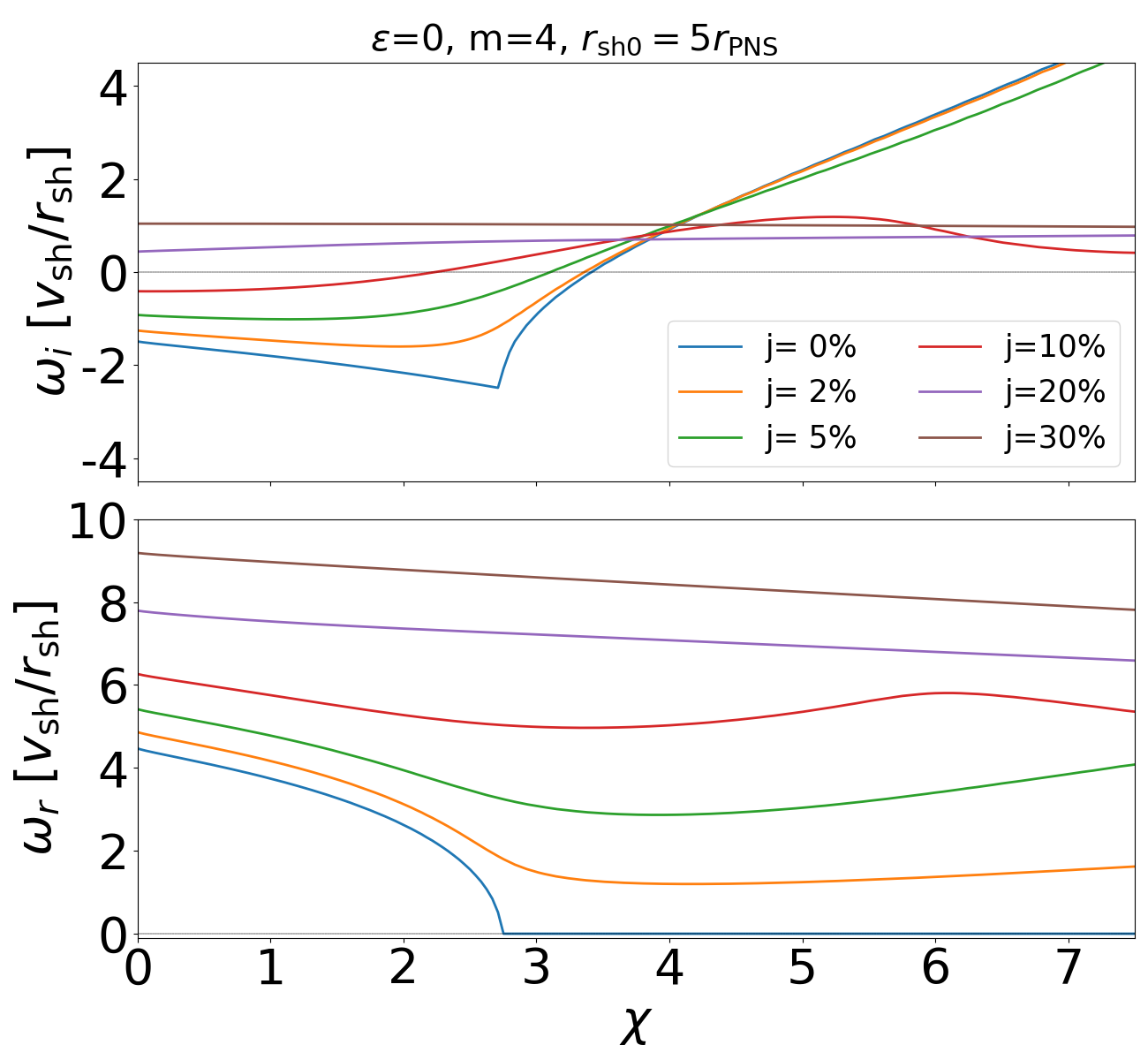}
    \caption{Evolution as a function of $\chi$ of the growth rate $\omega_i$ (top) and the frequency $\omega_r$ (bottom) of the fundamental mode with an azimuthal number $m=4$ for different values of the rotation rate.}
    \label{wi(chi)_rot}
\end{figure}

%----Growth rate
We consider in Fig.~\ref{wi(chi)_rot} the effect of the convection parameter on the eigenfrequency of the mode $m=4$ for different rotation rates. The mode $m=4$ was chosen because it is the first mode exhibiting a convective instability when $j=0$.
The most striking feature regarding the oscillation frequency (lower plot in Fig.~\ref{wi(chi)_rot}), compared to Fig.~\ref{wi(chi)}, is the loss of a sharp transition between an oscillatory instability for modest neutrino-heating and a purely growing instability for strong enough neutrino-heating. This effect blurs the distinction between SASI and convection. This distinction becomes difficult for $j\gtrsim 0.1$ as will be discussed in the next section. In this section, we focus on the regime of slow rotation ($j\lesssim 0.1$), in which the two instabilities can still be distinguished. 

\begin{figure}[h]
    \centering
    \includegraphics[width=\hsize]{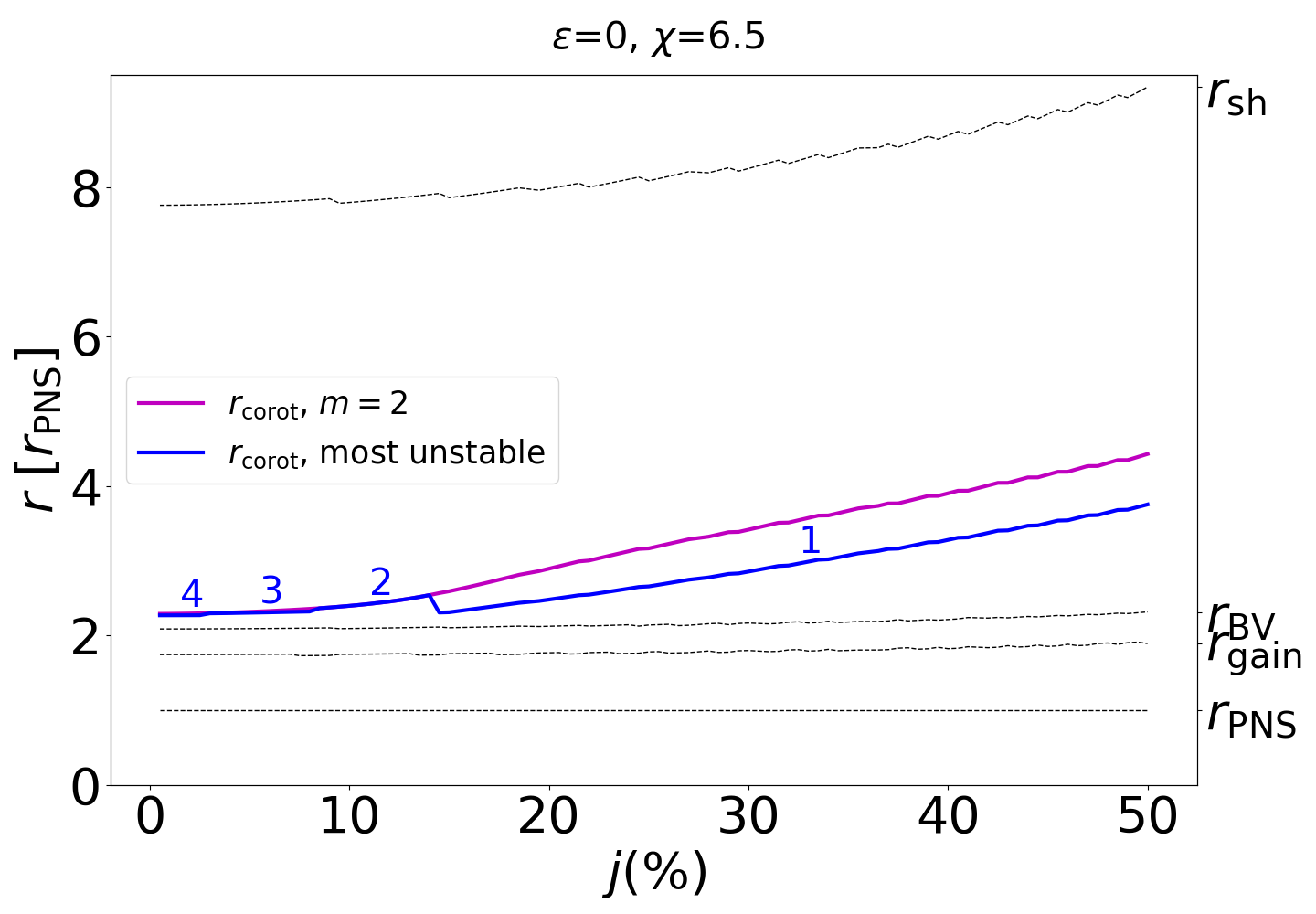}
    \includegraphics[width=\hsize]{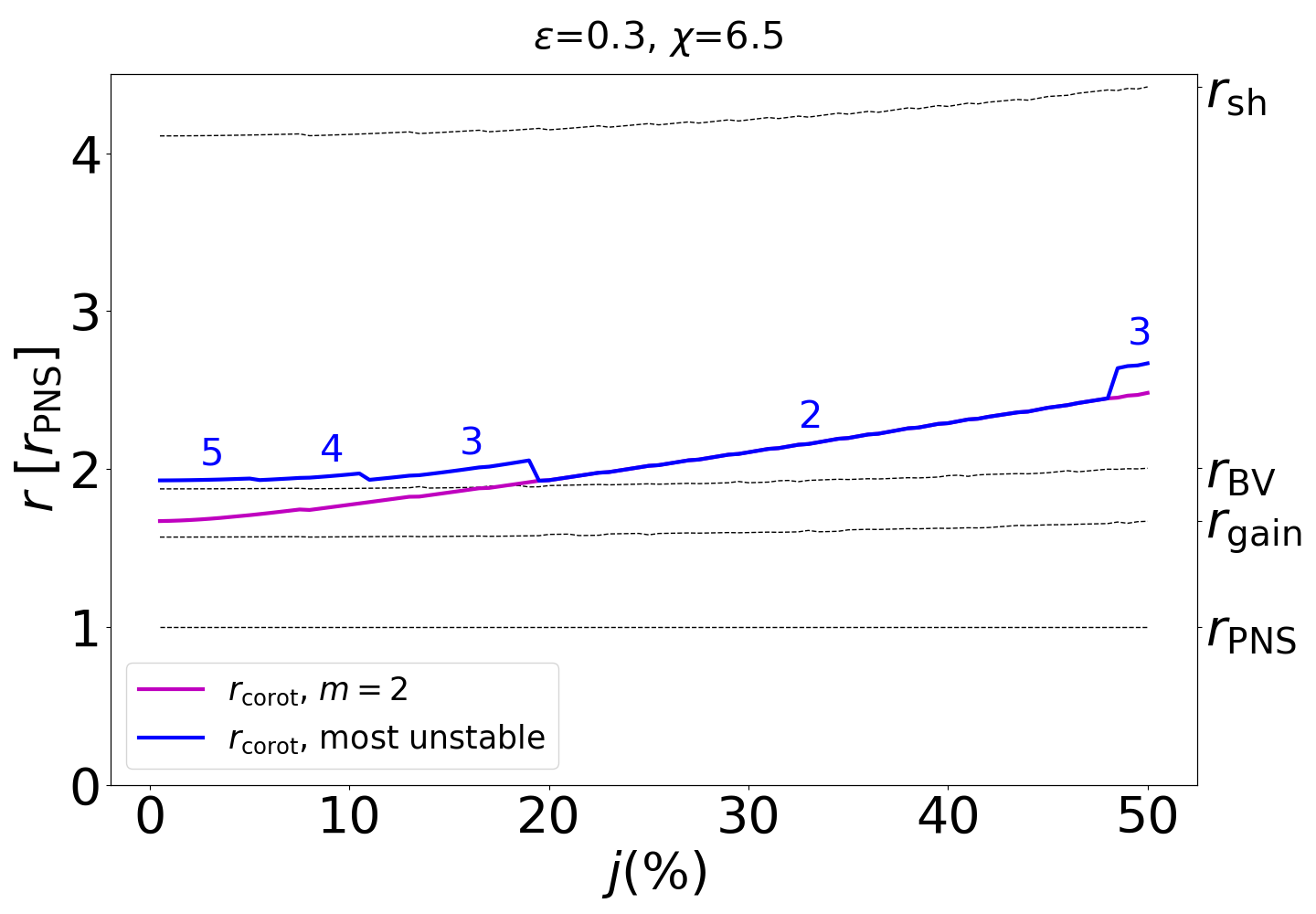}
    \caption{Evolution of the corotation radius $r_{\rm corot}$ as a function of the rotation rate for $\varepsilon=0$ (top) and $\varepsilon = 0.3$ (bottom). The corotation radius of the most unstable mode is represented with a blue line (numbers above the line specify the corresponding azimuthal number $m$). For comparison, grey lines show the PNS radius $r_{\rm PNS}$, the gain radius $r_{\rm gain}$, the radius of the Brunt-Väisälä frequency maximum $r_{\rm BV}$ and the shock radius $r_{\rm sh}$. With both values of the dissociation, the corotation radius of the most unstable mode is close to $r_{\rm BV}$ for slow rotation and moves away for higher rotation rates ($j>30\%$).}
    \label{rcorot_m2mdom}
\end{figure}

Even with modest angular momentum, the differential nature of rotation implies that convective motions are sheared, in addition to being entrained in rotation with a frequency intermediate between the shock and PNS rotational frequency. The region most unstable to buoyant motions, i.e. the radius $r_{\rm BV}$ corresponding to the largest Brunt-Väisälä frequency $N(r)$ (Eq.~\ref{definition_N}) is expected to contribute most to defining the phase frequency of the convective mode, i.e. the rotation frequency of the frame in which the convective pattern is stationary. In Fig.~\ref{rcorot_m2mdom} we compare $r_{\rm BV}$ with the corotation radius $r_{\rm corot}$ defined by Eq.~\eqref{definition_rcorot}. For small rotation rates ($j \lesssim 0.1$), we remark that $r_{\rm corot}$ is close to $r_{\rm BV}$ as anticipated for a purely convective mode.

\begin{figure}
    \centering
    \includegraphics[width=\columnwidth]{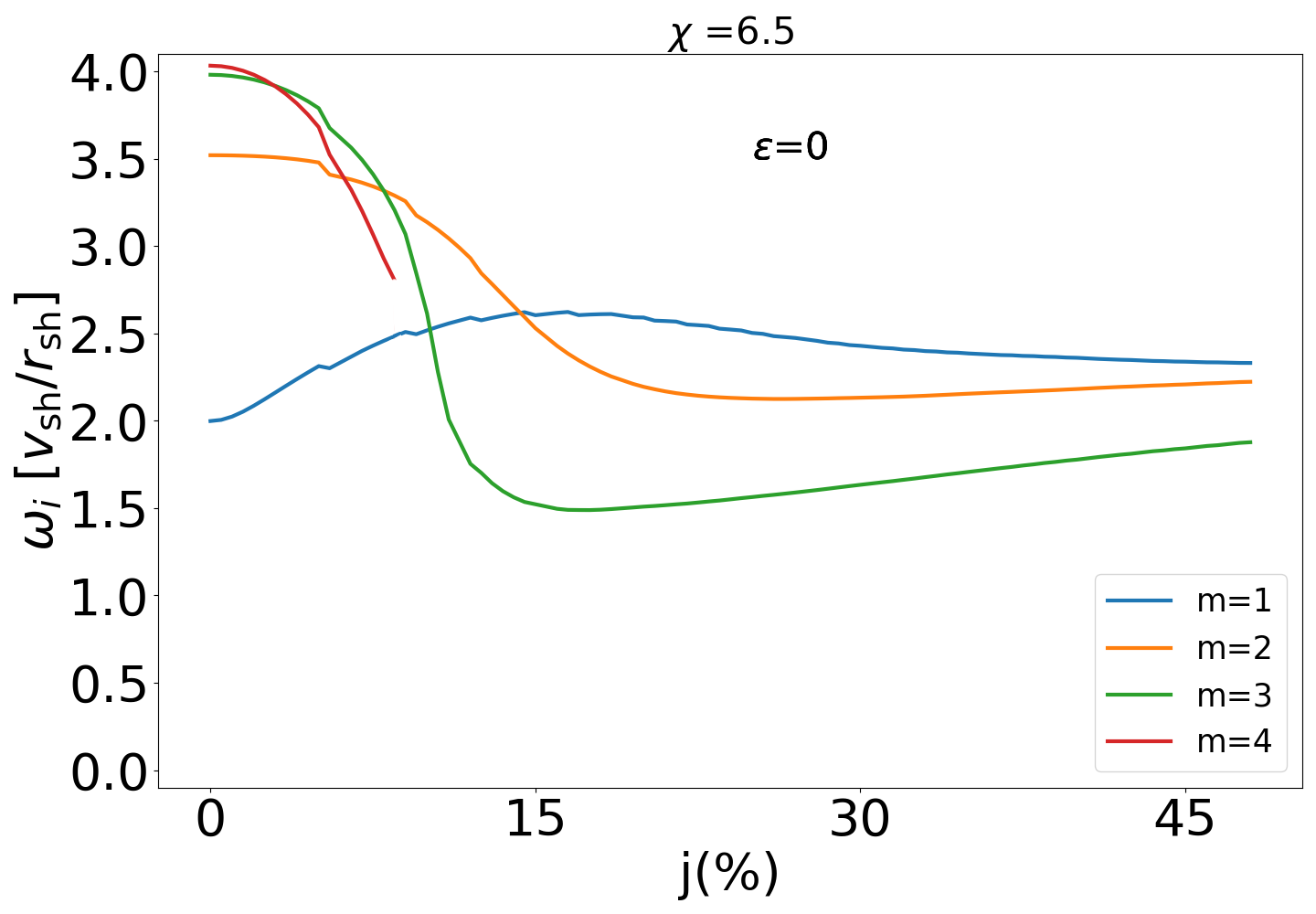}
    \includegraphics[width=\columnwidth,trim={0 0 0 1.5cm},clip]{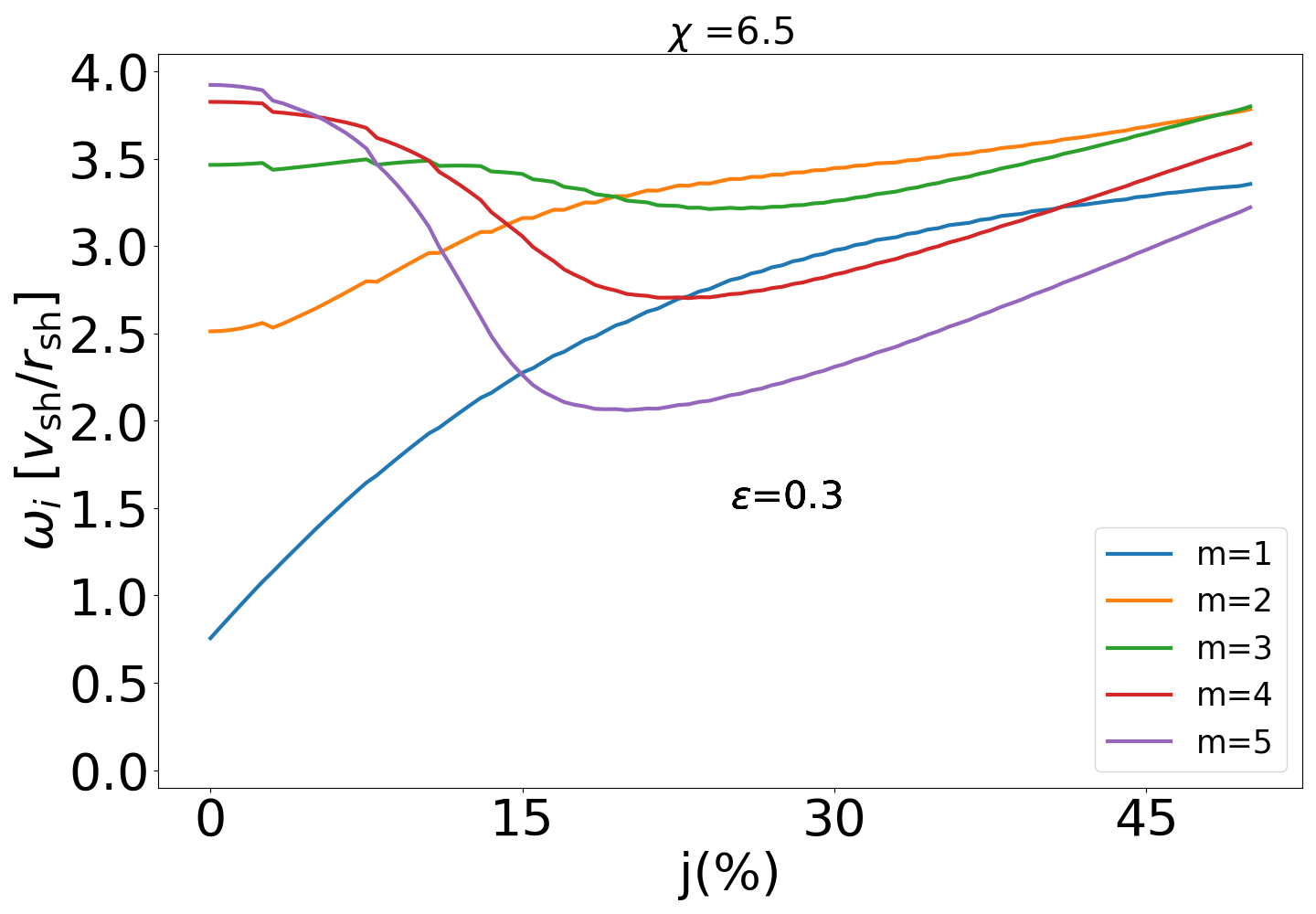}
    \caption{Growth rate evolution depending on the rotation rate for several modes at $\chi=6.5$ with $\varepsilon=0$ (top panel) and $\varepsilon=0.3$ (bottom panel). For $j=0$, $r_{\rm sh}(\varepsilon=0)=7.8r_{\rm PNS}$ and $r_{\rm sh}(\varepsilon=0.3)=4.1r_{\rm PNS}$. For this convection parameter, non axisymmetric convective modes are quenched by rotation. In the absence of rotation, all modes are above the transition from SASI to convection, except for the mode $m=1$ in the bottom panel.}
    \label{rot_effect_conv}
\end{figure}

Figure~\ref{rot_effect_conv} illustrates the effect of rotation on the growth rate of convective modes at $\chi=6.5$, which is significantly larger than the marginal stability threshold of the most unstable mode ($\chi=3.5-4$). For both dissociation rates considered ($\varepsilon=0$ and $\varepsilon=0.3$), the differential rotation reduces the growth rate of the convective instability, in particular for the smaller scale modes. The adverse effect on the convective instability is less abrupt for larger scale modes, which we interpret as a smaller effect of the shear between the shock and the PNS. As a result, larger angular scales become dominant for high rotation rates. At the largest scales ($m=1$ for $\varepsilon=0$, $m=1-2$ for $\varepsilon=0.3$), the effect of rotation is an increase of the growth rate which will be discussed in the next paragraph and section.

\begin{figure}
    \centering
    \includegraphics[width=\columnwidth]{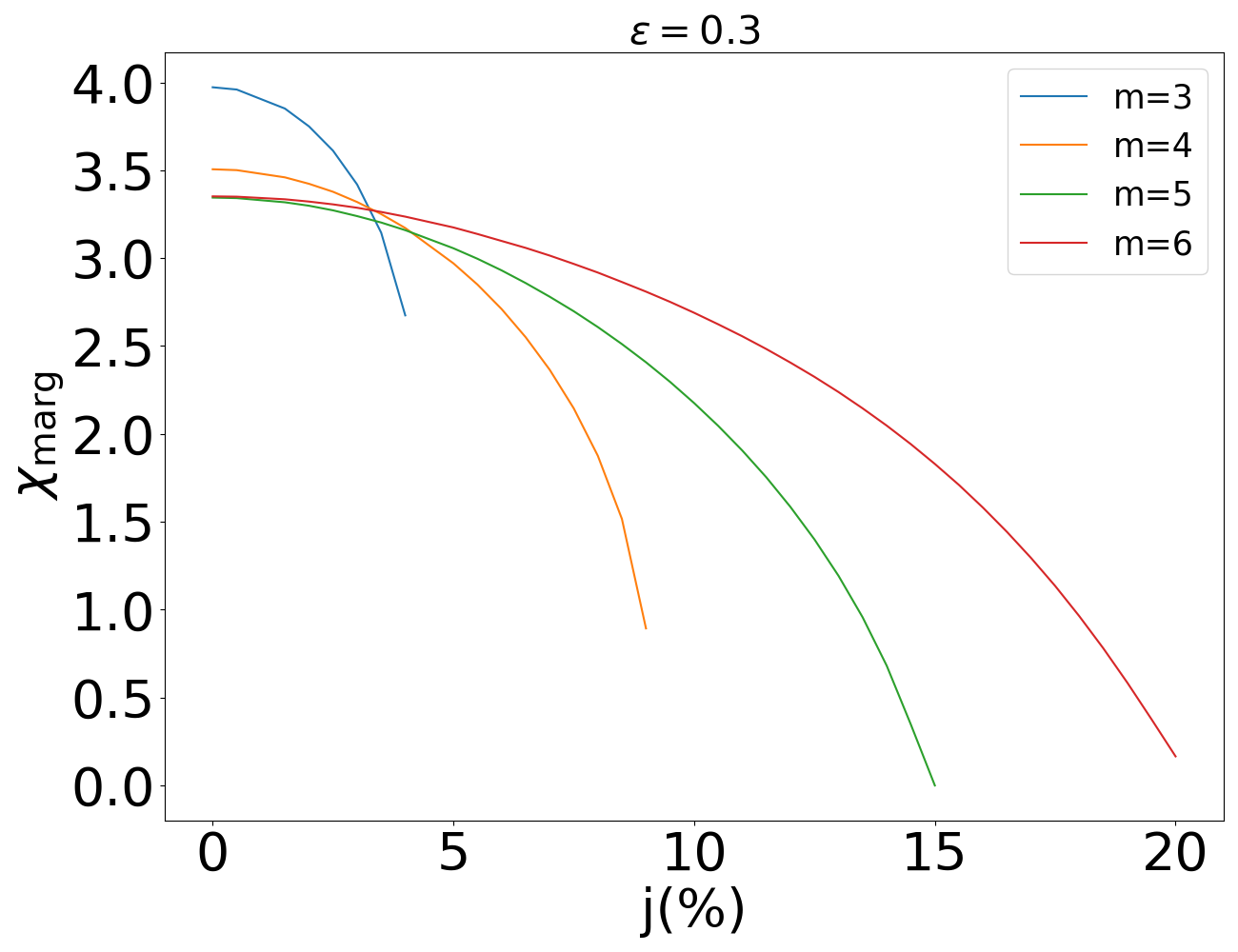}
    \caption{Impact of rotation on the convection parameter at marginal stability $\chi_{\rm marg}$. We consider several azimuthal numbers with $m>3$, for which SASI is stable in the absence of rotation. The $m=1$ and $m=2$  are not shown because they are unstable to SASI, which prevents a measure of $\chi_{\rm marg}$. The curves are cut when the mode is unstable for all convection parameters, in which case $\chi_{\rm marg}$ does not exist anymore. 
    }
    \label{chi_crit(j)_varm}
\end{figure}

Fig.~\ref{chi_crit(j)_varm} shows the effect of rotation on the convection parameter at marginal stability, referred to as $\chi_{\rm marg}$. This threshold for unstable convective modes decreases with rotation for all $m$ considered, thus showing a \emph{destabilising} effect on convection near its marginal stability. Such a destabilising effect is surprising and stands in striking contrast to the adverse effect of rotation at $\chi=6.5$ discussed in the last paragraph. We interpret this effect as a consequence of the mixed nature of the modes near the transition between SASI and convection. In this interpretation, the destabilising effect of rotation on convection is a residual effect of the rotational destabilisation of SASI which is well established numerically \citep{YamasakiFoglizzo2008,Blondin+2017,Walk+2022} but has never been explained physically.
%in the regime where the convective mode retains some characteristics of SASI near the transition between the two instabilities. 
The effect of rotation is thus expected to depend on how close the convection parameter is to the SASI/convection transition. This expectation is in agreement with Fig.~\ref{chi_crit(j)_varm}, which shows that the convection parameter at marginal stability $\chi_{\rm marg}$ decreases faster with rotation for modes with smaller $m$. In the absence of rotation, when $m$ increases, the transitional $\chi_{\rm trans}$ from SASI to convection decreases (see Fig.~\ref{wi(chi)}) while the marginal stability $\chi_{\rm marg}$ depends little on the azimuthal number. As a result, the difference $\Delta\chi \equiv \chi_{\rm trans}-\chi_{\rm marg}$ increases with $m$, and the residual effect of SASI decreases. This highlights the mixed SASI/convection nature of the modes near $\chi_{\rm trans}$ for small rotation rates.

\subsubsection{Mixed SASI-convective mode induced by rotation}
\label{sec:mix_modes}

For intermediate rotation rates ($0.1<j<0.3$), the behaviour of the modes is modified by rotation to such an extent that it becomes impossible to distinguish clearly between convective and SASI modes. As described in this section, the modes in this regime retain characteristics of both instabilities and are best understood as mixed SASI/convection/rotation modes. 

\begin{figure*}[h]
    \centering
    \includegraphics[width=\hsize]{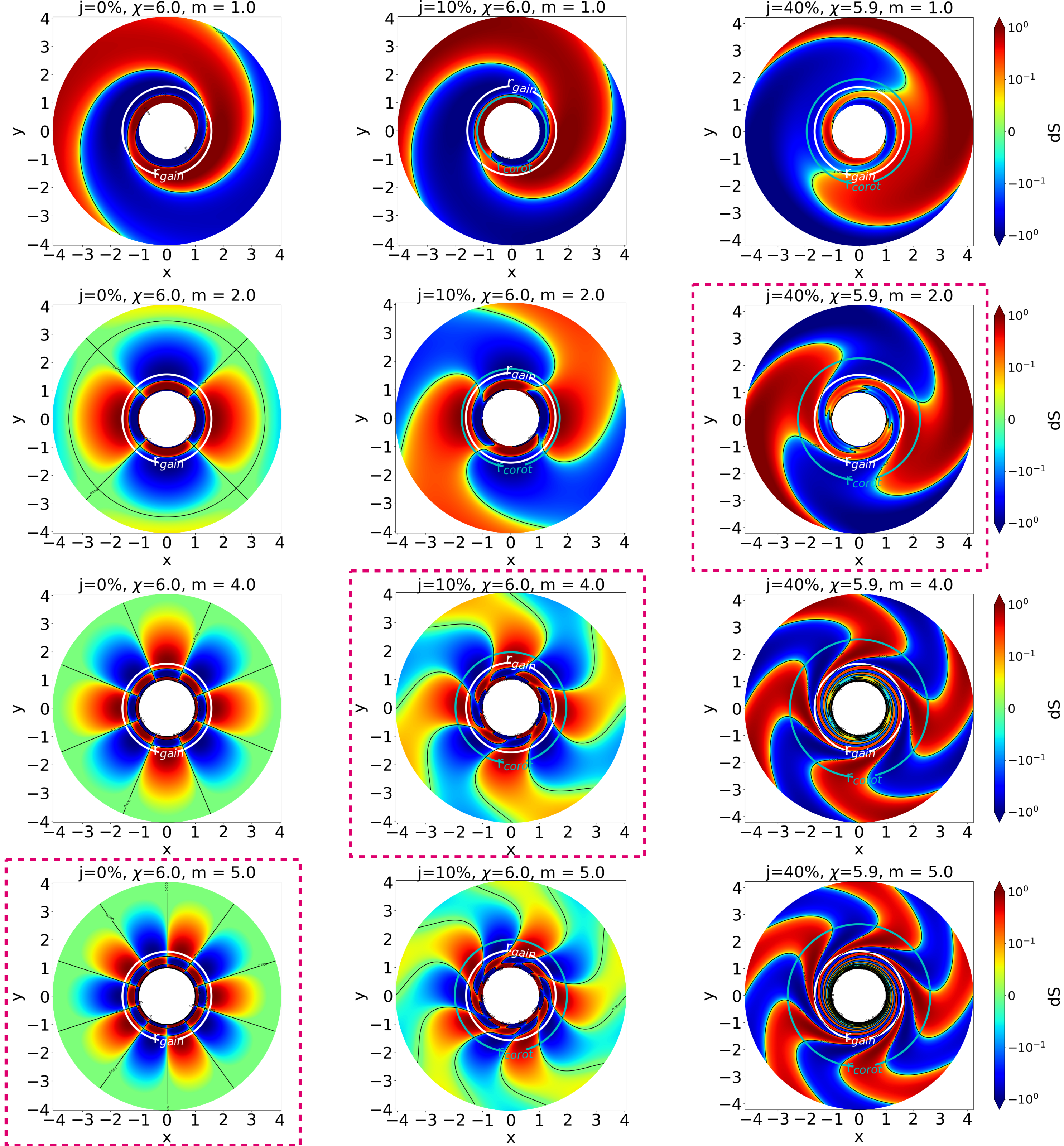}
    \caption{Spatial structure of the eigenmode entropy perturbation in the equatorial plane, for $\varepsilon=0.3$ and $\chi\sim6$. Each row corresponds to a value of $m=(1,2,4,5)$ (from top to bottom) and each column to a rotation rate with $j=(0,0.1,0.4)$ (from left to right). In magenta and cyan, we represented the gain and corotation radii, respectively. The diagonal sequence of framed plots corresponds to the most unstable modes. Coordinates $x$ and $y$ are expressed in $r_{\rm PNS}$ unit. The deformation induced by rotation is stronger for smaller-scale modes.}
    \label{structure}
\end{figure*}

%Jolies images
In order to gain a deeper understanding of the transition between SASI dominated and buoyancy dominated regimes including rotation, we show in Fig.~\ref{structure} the entropy structure of the eigenfunction for the modes $m=1,2,4,5$ for three values of the angular momentum $j=0$, $0.1$ and $0.4$ and $\chi\sim 6$. Without rotation (first column in Fig.~\ref{structure}), the entropy structure of SASI (first line) is uniformly spread in the radial direction as a spiral pattern. It is clearly distinct from the structure of convection, whose perturbations are more localised in the vicinity of the gain radius and do not display any spiral pattern. In the second column of this figure, we note that the $m=1$ spiral SASI pattern is little affected by a moderate rotation with $j=0.1$ (compare left and middle columns). This can be interpreted by the fact that for such moderate rotation, the corotation radius is close to the PNS surface. For the same rotation rate, the differential rotation shears the convective cells with $m=2-5$. This leads to spiral structures that are smaller but similar to a SASI pattern. However, the convective nature of the mode is still visible by the localisation of the perturbations near the gain radius. These similar structures suggest that the mode is not purely convective, but a mixed SASI/convection/rotation regime.

\begin{figure*}[htbp]
    \centering
    \includegraphics[width=0.72\textwidth]{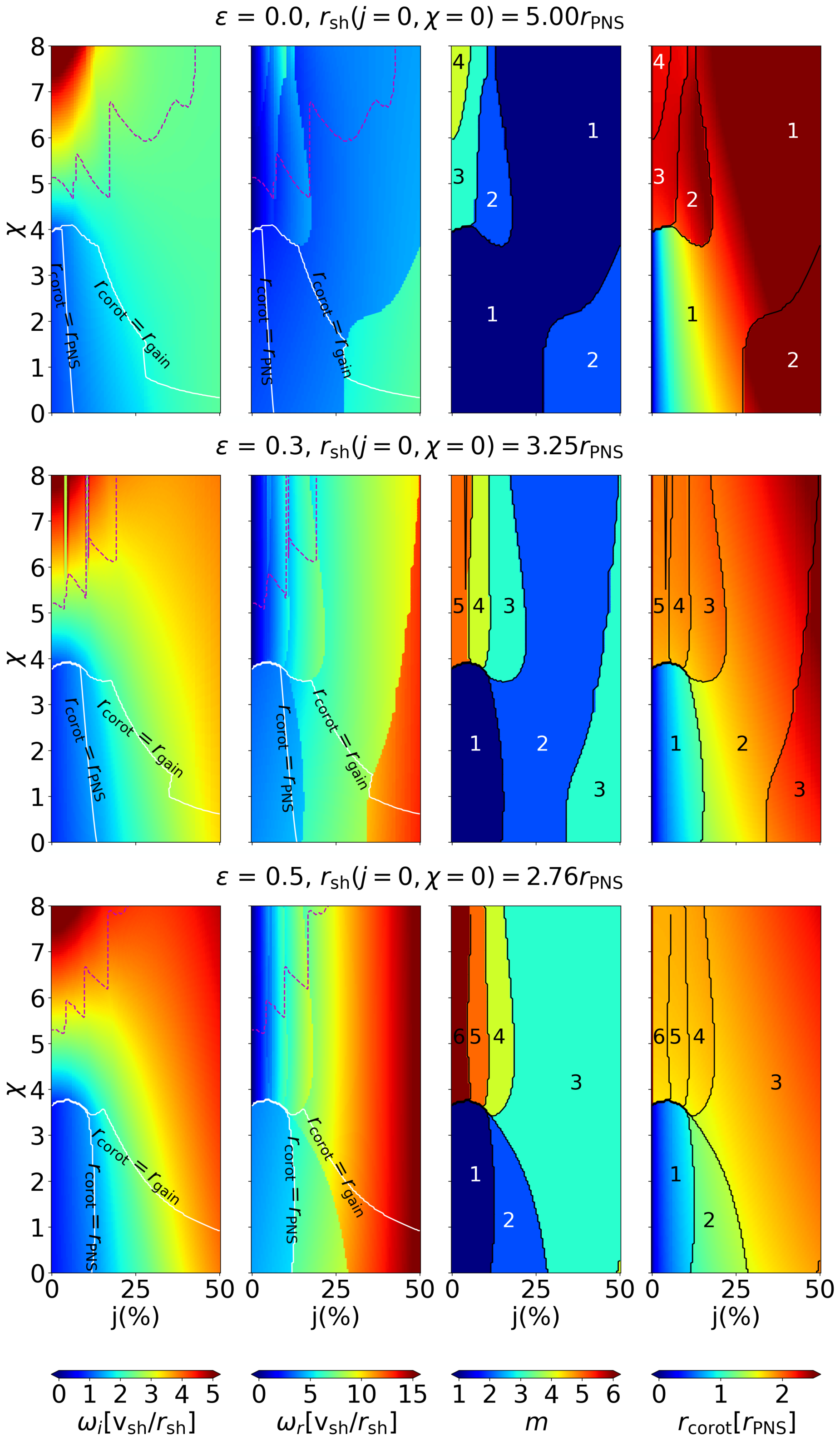}
    \caption{From left to right, maps of the growth rate $\omega_i$, frequency $\omega_r$, azimuthal number $m$ and corotation radius of the most unstable fundamental mode with three dissociation rates $\varepsilon=0$ (top row), $0.3$ (middle row) and $0.5$ (bottom row). Above the magenta dashed line, the growth rate of equatorial perturbations decreases with rotation. In the two left columns, the white lines show where the corotation radius equals $r_{\rm PNS}$ and $r_{\rm gain}$. In the two right columns, the black lines highlight the variation of the dominant azimuthal number $m$. The numbers in the third and fourth columns highlight the azimuthal number of the most unstable mode.}
    \label{pano}
\end{figure*}

A global panorama of the parameter space is shown in Fig.~\ref{pano} where $\chi$ and $j$ are varied for three values of the dissociation parameter $\varepsilon=0$, $0.3$ and $0.5$ (upper to lower rows). The four columns represent the growth rate and oscillation frequency of the most unstable mode, its azimuthal number $m$ and the corotation radius $r_{\rm corot}$. A prominent effect of dissociation is a reduction of $r_{\rm sh}$ and thus the timescale of advection across the gain region.

%variation frequence
For small values of $j$, a sharp jump in frequency and $m$ is visible in Fig.~\ref{pano} where neutrino heating exceeds the threshold $\chi\sim 4$, corresponding to the transition from SASI to the convective instability as the most unstable mode. However, as rotation increases, the frequency of the convective instability increases until it becomes similar to the SASI frequency. When the angular momentum reaches $j_{\rm crit}\sim 0.08-0.12$, the jump disappears and it becomes impossible to distinguish clearly convective from SASI modes: this is the regime of mixed modes.

%corotation
The occurence of mixed modes can also be recognised in the evolution of the corotation radius shown in Fig.~\ref{rcorot_m2mdom}.
Above a critical rotation rate, the corotation radius indeed moves away from the Brunt-V\"ais\"al\"a radius, where it is located for a convective mode. We remark that $r_{\rm corot}$ shifts outwards for rotation rates exceeding $j\sim10\%$, indicating a lower frequency than expected for a purely convective mode. This slower oscillation frequency compared to the rotation rate of the most buoyant layer indicates that some physical process associated to rotation and/or SASI contributes significantly to the instability mechanism.
However, the decrease of the azimuthal number of the dominant mode (see below) maintains the corotation close to the Brunt-Väisälä radius, indicating that convection still plays a role in the instability mechanism. 

%variation du $m$ dominant
We remark in the 3rd column of Fig.~\ref{pano} that the azimuthal number $m$ of the most unstable mode decreases with the rotation rate for $\chi>4$, while it increases for $\chi<4$. The change of dominant convective scale induced by rotation is not predicted by Eq.~\eqref{eq_m}: comparing the magenta and blue triangles in Fig.~\ref{m_dom(eps)} indicates a shift to larger angular scales induced by a moderate angular momentum $j=0.1$, despite a negligible change in the size of the gain region. For this small rotation rate, the shortcoming of Eq.~\eqref{eq_m} is not surprising when considering the rotation-induced distortion of the convective cells in Fig.~\ref{structure} and the residual influence of SASI. The rotational stabilisation of perturbations with a large azimuthal wave number $m$ can be understood as a consequence of their strong shear (Eq.~\ref{doppler_omega}) which induces a short radial wavelength $2\pi v/\omega'$ for advected perturbations near the PNS, and thus a phase mixing particularly visible in the lower right corner of Fig.~\ref{structure}. 

\begin{figure}[h]
    \centering
    \includegraphics[width=\hsize]{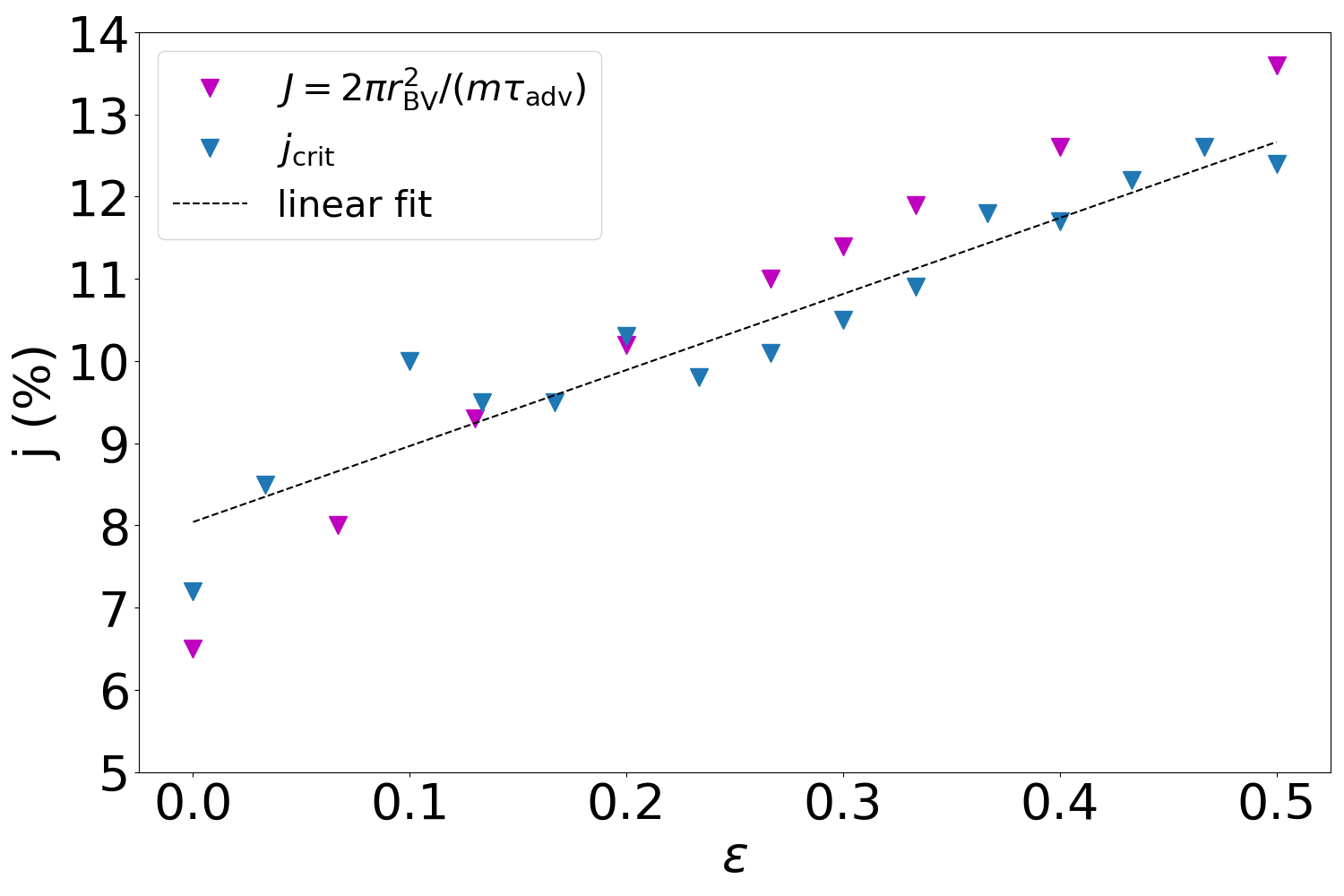}
    \caption{Evolution of the critical rotation rate $j_{\rm crit}$ depending on the dissociation using two methods. This critical rotation corresponds to the apparition of the mixed-modes. The higher the dissociation, the later the transition. For the blue points, we used the discontinuity end of the dominant m at $\chi\sim4$. For the magenta points, we used the advection timescale $\tau_{\rm adv}$ from the shock to the gain radius for $m=4$ and $\chi\sim4$.}
    \label{jcrit(eps)}
\end{figure}

The third column of Fig.~\ref{pano} can be used to better constrain the domain of existence of these mixed modes. For small rotation rates, the azimuthal number contours converge to a line at $\chi\sim 4$. It corresponds to the transition from purely SASI modes to purely convective modes, across which the azimuthal number and frequency of the dominant mode is discontinuous. Above this boundary, the high $m$ modes dominate, and their frequency depends linearly on the rotation rate. However, this frontier disappears for higher rotation rates, and the transition from small $m$ to higher $m$ is continuous. The value of the rotation rate $j_{\rm crit}$ when the frontier disappears slightly increases when the dissociation rate increases (blue points in Fig.~\ref{jcrit(eps)}). This behaviour can be quantitatively reproduced if we assume that mixed modes appear when the frequency of the convective mode $\simeq m\Omega(r_{\rm BV})$ matches the frequency of the corresponding SASI mode in the absence of rotation or heating $\simeq 2\pi/\tau_{\rm adv}$. This criterion defines a critical specific angular momentum 
\begin{equation}
    J_{\rm mix} \equiv \frac{2\pi r_{\rm BV}^2}{m\tau_{\rm adv}},
\end{equation}
which reproduces remarkably well the variation of $j_{\rm crit}$ when considering the mode $m=4$ (magenta points in Fig.~\ref{jcrit(eps)}). This is the signature of the growth of mixed modes SASI/convection/rotation.

%-----Growth rate mixed modes

The first column in Fig.~\ref{pano} shows the adverse effect of rotation on the growth rate of the convective instability for $\chi>5$, as discussed in Sect.~\ref{sec:convection_rotation}. For each rotation rate, the magenta line highlights the value of $\chi$ above which rotation has an adverse effect on the growth of convection in the equatorial plane. In this region, the dynamic of the flow is dominated by axisymmetric convective modes in the plane $(r,z)$, that are expected to be little affected by rotation for $j\in[0,0.3]$. This stands in contrast with the beneficial effect of rotation on the growth rate of SASI for $\chi<3-4$, as discussed in Sect.~\ref{sec:sasi_rotation}, where spiral modes are therefore expected to dominate the dynamics.
 Interestingly, rotation has a beneficial effect on the growth rate of the dominant non-axisymmetric convective modes for moderate convection parameters $\chi \simeq 4-5$ (Fig.~\ref{pano}). The same behaviour can be observed for the $m=4$ mode in the upper plot of Fig.~\ref{wi(chi)_rot}, where the growth rate for $\chi\sim 3-4$ is significantly enhanced by rotation. It is also linked to the easier onset of convection with rotation shown by the decrease of $\chi_{\rm marg}$ in Fig.~\ref{chi_crit(j)_varm}, as discussed in Sect.~\ref{sec:convection_rotation}. All these observations suggest that the perturbation of the shock induced by the convective instability produces an advective-acoustic cycle favourable to enhance the convective instability, as already suggested by the entropy structure in Fig.~\ref{structure}. 
The mode $m=1$ in Fig.~\ref{rot_effect_conv} calls for a particular attention, with a significant enhancement of the growth rate by rotation  (this is also true for the $m=2$ mode in the lower panel of the same figure) in a regime $\chi=6.5$ a priori dominated by convection with an adverse effect of rotation. This peculiar behaviour of the $m=1-2$ modes can also be explained by a mixed state of this mode, with the advective-acoustic cycle interacting constructively with the convective mode. Indeed, for these low values of $m$ the transition from SASI to convection takes place at a rather large value of $\chi$ (e.g. $\chi \simeq 6$ for $m=1$ in Fig.~\ref{wi(chi)}). As a consequence, mixed modes with a beneficial effect of rotation are expected for larger values of $\chi$ close to this transition.

\begin{figure}[h]
    \centering
    \includegraphics[width=\columnwidth]{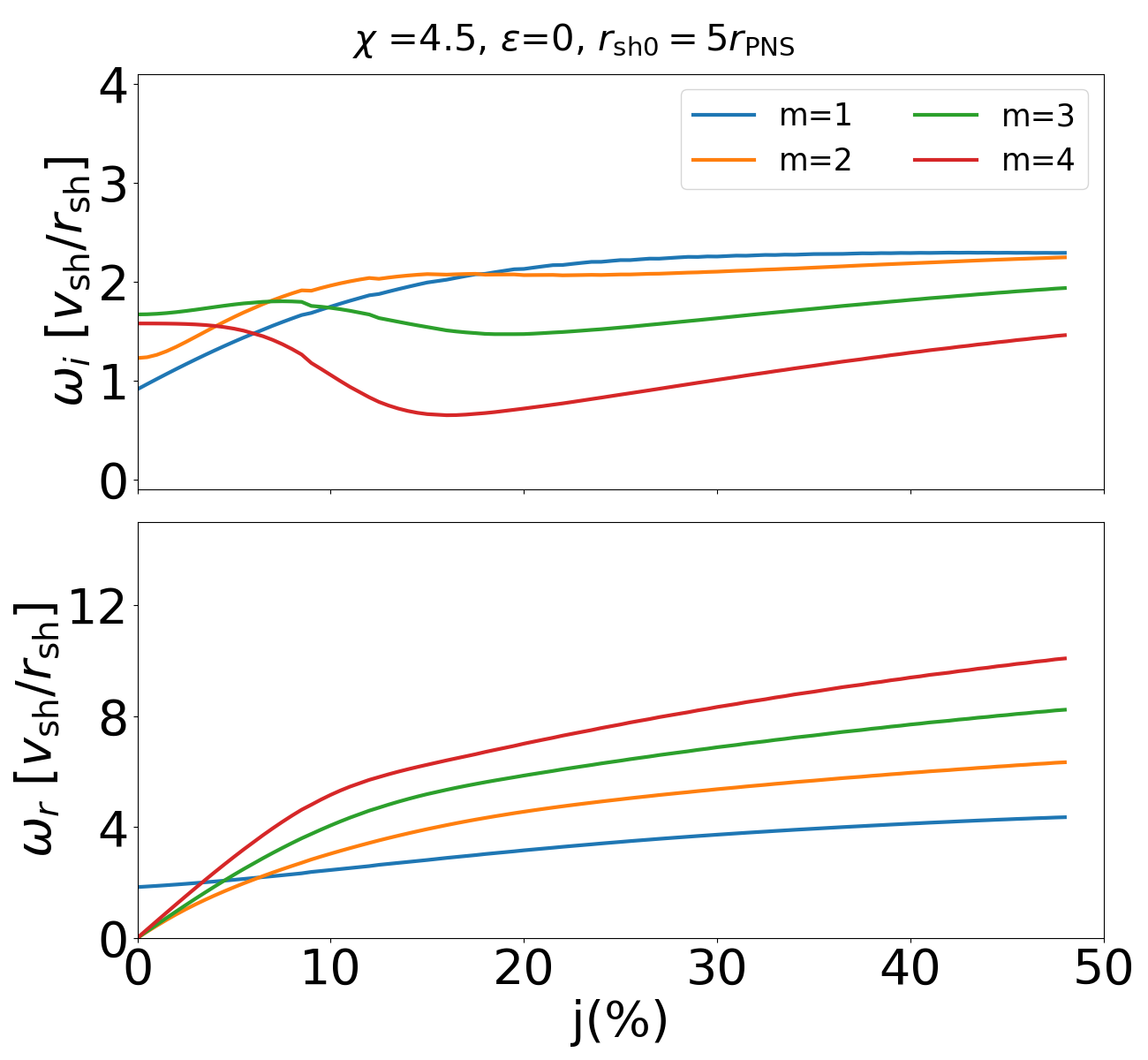}
    \caption{Growth rate (top) and frequency (bottom) evolution of several modes depending on the rotation rate $j$, for $\chi=4.5$, $\varepsilon=0$ and $r_{\rm sh}(j=0)=6.8r_{\rm PNS}$.}
    \label{evol_wi}
\end{figure}

\begin{figure}[h]
    \centering
    \includegraphics[width=\columnwidth]{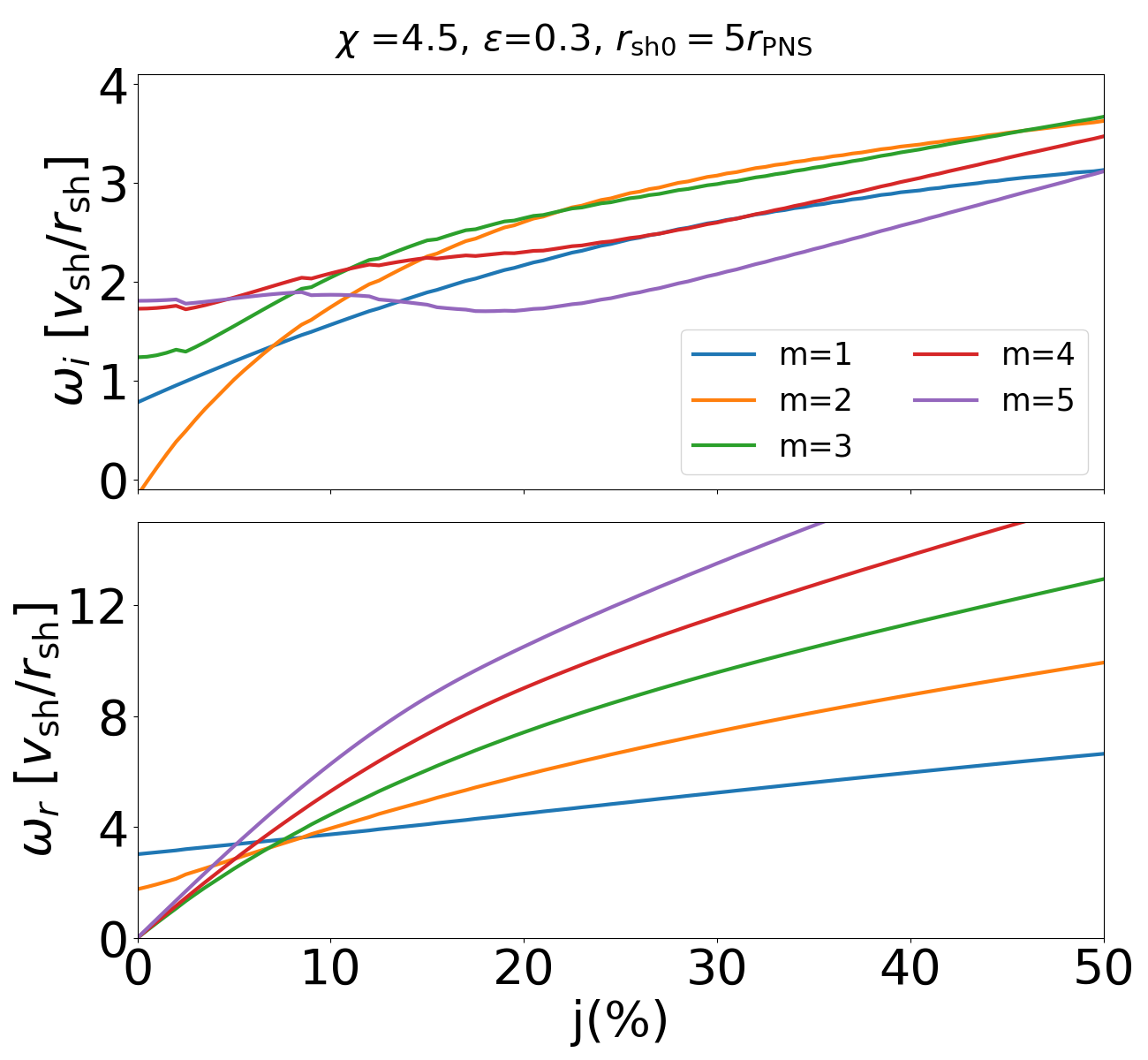}
    \caption{Same as Fig.~\ref{evol_wi} but for $\varepsilon=0.3$ and $r_{\rm sh}(j=0)=3.9r_{\rm PNS}$.}
    \label{evol_wi_dis}
\end{figure}

Figs.~\ref{evol_wi} and \ref{evol_wi_dis} illustrate the effect of rotation on both instabilities in two configurations mainly differing by their shock radius, induced by different values of the dissociation parameter. The value $\chi=4.5$ is chosen in the region where the most unstable mode is convective and the growth rate increases with rotation (below the magenta line of the first column in Fig.~\ref{pano}). 
In the bottom panel of these figures, the oscillation frequency without rotation ($j=0$) reveals the convective ($\omega_r=0$) or SASI ($\omega_r\neq0$) nature of the modes. We can see that the convective frequency, enhanced by rotation, is similar to the SASI frequency for $j\in[0.02,0.08]$. The top panels of this figure show that the effect of rotation varies depending on the nature of the instability. The growth rate of SASI modes is enhanced by rotation, as in \cite{YamasakiFoglizzo2008} and \cite{Blondin+2017}. As mentionned in Sect.~\ref{sec:convection_rotation}, for convective modes, the effect of rotation depends on the proximity of $\chi$ to the transition value $\chi_{\rm trans}$ (defined in Sect.\ref{sec:sasi_rotation}).% defining the transition of a mode between SASI and convection. Defining $\Delta \chi\equiv\chi-\chi_{\rm trans}$, 
 The rotation can induce an increase of the growth rate, similar to the effect of rotation on SASI, if $\Delta \chi<1$. Otherwise, the convective modes are quenched by rotation, as observed in Fig.~\ref{rot_effect_conv}. 
%Comparing Figs.~\ref{evol_wi} and \ref{evol_wi_dis}, the increase of the frequency and growth rate induced by rotation is stronger for $\varepsilon=0.3$.

\begin{figure}
    \centering
    \includegraphics[width=\columnwidth]{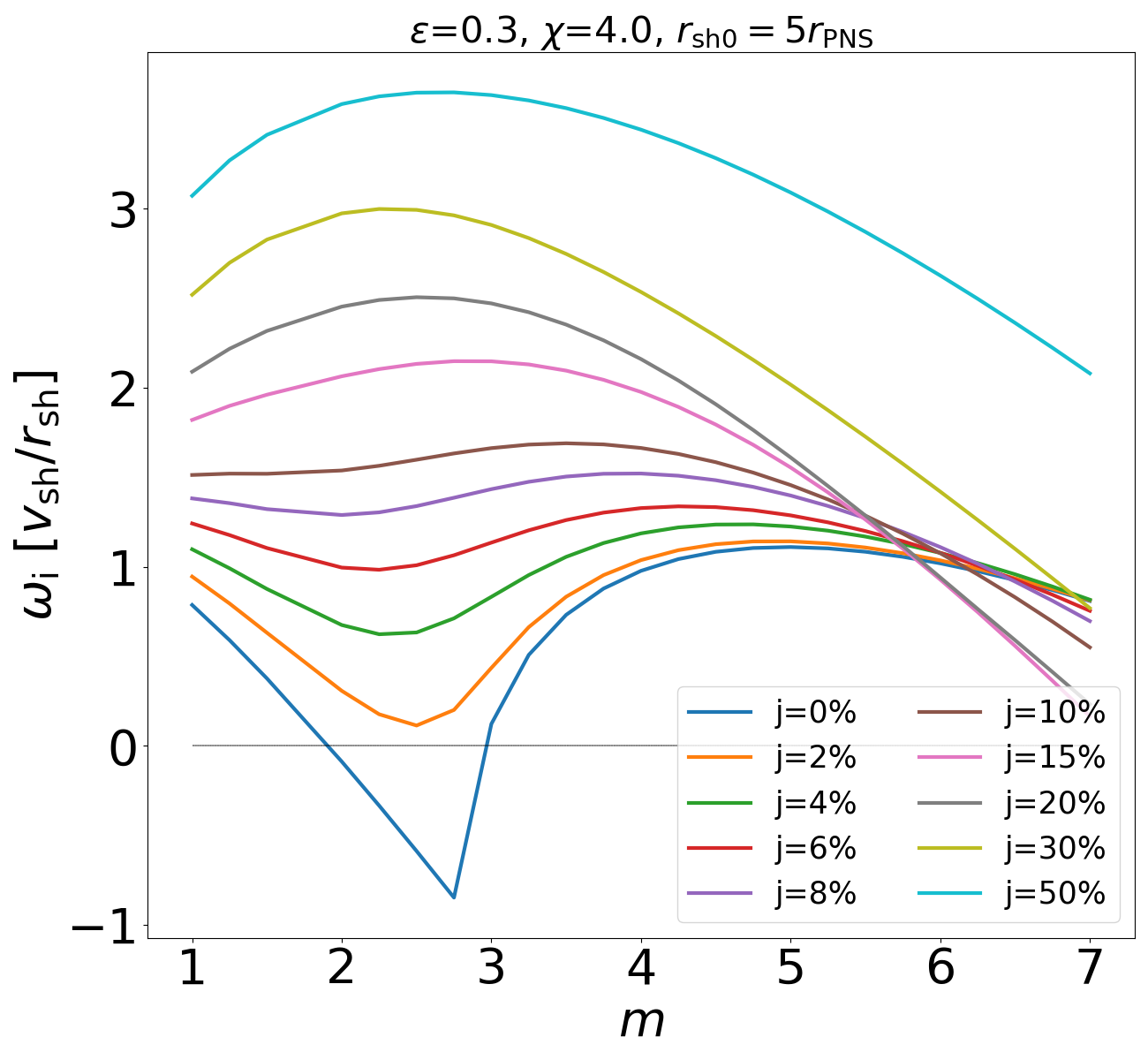}
    \caption{Evolution of the growth rate depending on the azimuthal number $m$ for several rotation rates and $\chi=4.5$.}
    \label{wi(m)_varrot}
\end{figure}

Figure~\ref{wi(m)_varrot} illustrates the global complexity of the impact of rotation depending on the azimuthal number. The growth rate of SASI modes ($m=1,2$) is increased as seen on the left part of the plot. However, the effect of rotation on convective modes depends on both the rotation and the azimuthal number. For $m\sim3$, the destabilising effect of rotation on the convective instability can be clearly seen in continuity with the increase of the SASI growth rate. For $m \in [3,5]$, the effect of rotation on convection is similar to its effect on SASI, but with a smaller magnitude. 
For higher $m\gtrsim6$ and low rotation rates, rotation has an adverse effect and leads to a decrease of the growth rate. The effect changes for high rotation rates and will be discussed in Sect.~\ref{part:high_rot}.

In summary, rotation smoothens the transition from SASI to convection, leading to a mixed mode regime. These mixed modes appear when the corotation radius moves away from the Brunt-Väisälä radius and exist up to $j\sim30\%$ where the properties of the modes start to change to the regime induced by strong rotation.

\subsection{Strong rotation-induced instability}
\label{part:high_rot}

The mixed state of the modes seems to disappear when the rotation becomes too strong.
When the rotation rate exceeds $j\sim20\%$, Fig.~\ref{wi(chi)_rot} shows that both the frequency and the growth rate of the $m=4$ mode, when expressed in units of $v_{\rm sh}/r_{\rm sh}$, become strikingly insensitive with respect to the convection parameter. Rotation also changes the behaviour of the most unstable mode $m$ as a function of heating. While the dominant $m$ increases with $\chi$ for small and intermediate rotations, it decreases slightly with heating for higher rotation rates. Rotation also diminishes the influence of the parameter $\chi$ on the frequency and growth rate of the most unstable mode for $j\gtrsim0.3$ (Fig.~\ref{pano}). This suggests that the mode loses its convective nature and that the buoyancy driven by the entropy gradient no longer plays an important role in the strong rotation regime.
Another indication 
%that the role of buoyancy in the instability mechanism has been erased by rotation 
comes from the fact that the corotation radius moves definitively away from the Brunt-Väisälä radius for high rotation rate $j\gtrsim30\%$ (Fig.~\ref{rcorot_m2mdom}). 

The right column of Fig.~\ref{structure} illustrates the structure of the modes in the strong rotation regime. The localisation around the Brunt-Väisälä radius, a characteristic of the convective instability that was still visible at intermediate rotation, is absent in this regime. The radial profile of entropy perturbations suggests that it is driven by the deformations of the shock rather than by the region of maximum buoyancy. The spiral mode structure of SASI is also clearly perturbed by rotation due to the corotation radius, which is well inside the domain. The pattern is similar to the one obtained by \cite{Blondin+2017} at high rotation rates without heating. In an adiabatic flow where entropy perturbations are simply advected, the entropy pattern would trace the flow lines and display a change of direction at the corotation radius since the pattern rotates faster than the outer flow and slower than the inner flow. In Fig.~\ref{structure}, a change in the direction of the spiral pattern is clearly visible in the strong rotation case (right column). The radial location of the pattern extrema is slightly above the corotation radius due to the non-adiabatic heating/cooling functions. 

For these high rotation rates ($j>0.3$), it seems that the modes are not mixed SASI/convection/rotation any more, but result from a rotation-induced instability. Some characteristics of this instability can be inferred from the evolution of the growth rate. Comparing Figs.~\ref{evol_wi} and \ref{evol_wi_dis}, we remark that when the modes properties do not depend on the $\chi$ parameter, the dependence of the modes on the rotation rate is stronger for increasing dissociation rates. In addition, the slope of this increase is approximately independent of the value of the azimuthal number. This behaviour of the mode for high rotation rates can also be seen in Fig.~\ref{wi(m)_varrot}. We notice that the rotation enhances the growth rate of every mode. In particular, for initially convective modes, this effect is opposite to the rotation effect for low rotation rates.

\section{Expected consequences on the gravitational wave signature\label{sec_GW}}

The mode frequencies computed with a linear analysis of the PNS have been shown to reproduce well the frequencies present in the gravitational waves spectra in non-linear numerical simulations \citep{TorresForne+2018,TorresForne+2019,westernacher-schneider19,westernacher-schneider20}. Our work can thus be used to predict the frequency of gravitational wave emission.
We first focus on the properties of the fundamental mode $m=2$ which has a direct impact on the production of gravitational waves according to the quadrupole formula.
We include the effect of dissociation with $\varepsilon\sim0.3$ %-0.5$
\citep{Fernandez+2009,Fernandez+2014,Huete+2018} and consider the typical range of shock radius  $r_{\rm sh}\sim (3-7) r_{\rm PNS}$. 

%-----Corotation radius
Figure~\ref{rcorot_m2mdom} (bottom) shows that the mode $m=2$ is dominant in the high rotation regime $j\in\left[0.2,0.5\right]$ for $\chi=6.5$ and $\varepsilon=0.3$. 
%This result remains true when the heating rate varies within this range of rotation rates. 
The modes $m=2,3$ are actually dominant over a large domain of the parameter space ($j, \chi$) according to the third column in Fig.~\ref{pano} for $\varepsilon=0.3-0.5$.

%---Frequency m=2

\begin{figure}[h]
    \centering
    \includegraphics[width=\hsize]{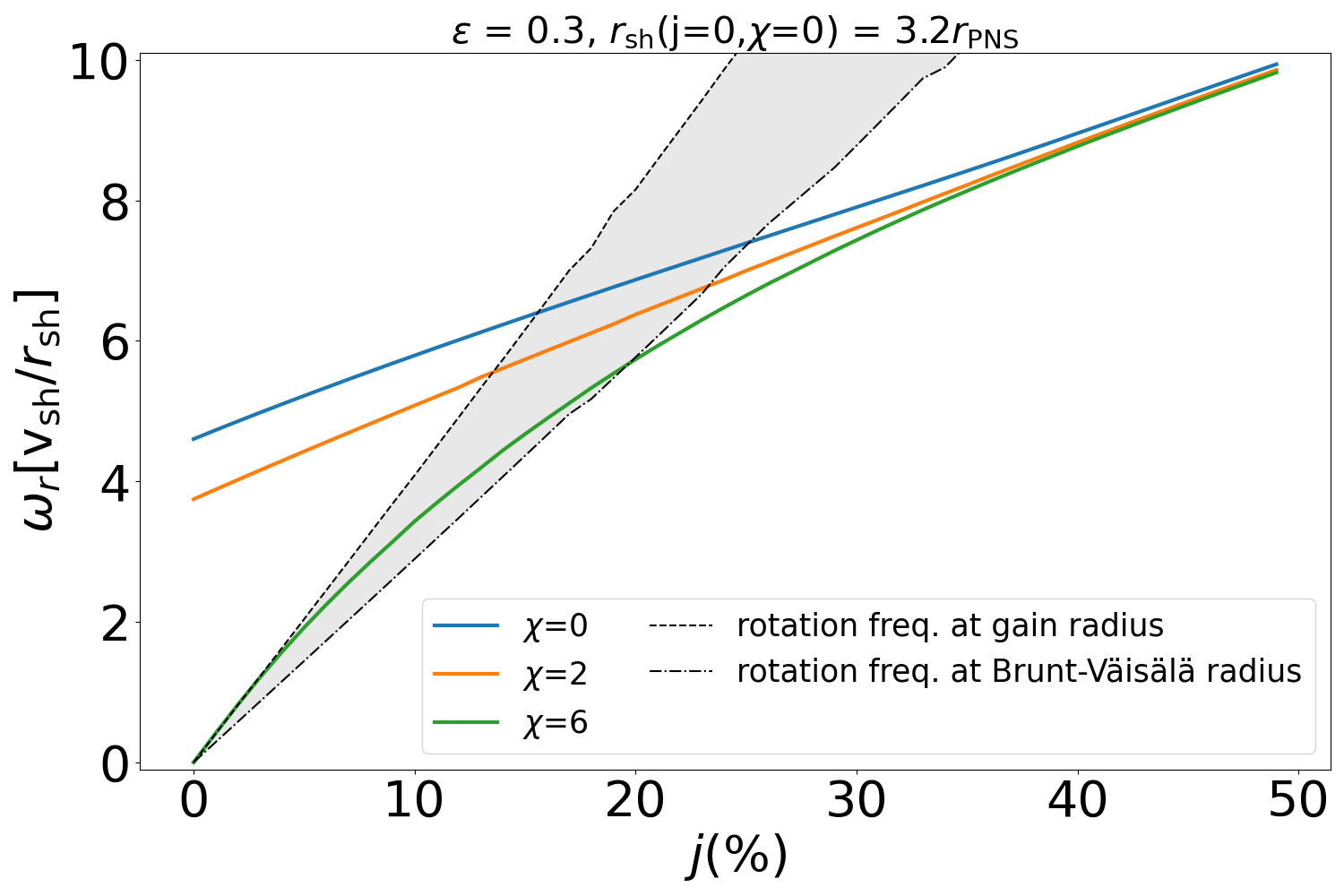}
    \caption{Frequency of the $m=2$ fundamental mode as a function of the specific angular momentum $j$ for several values of $\chi$. The grey-shaded region illustrates the frequency range corresponding to a corotation radius located between the gain radius $r_{\rm gain}$ and the radius $r_{\rm BV}$ maximising the Brunt-V\"ais\"al\"a frequency. The frequency is expressed in units of the approximate inverse advection timescale $v_\mathrm{sh}/r_\mathrm{sh}$.}
    \label{wr(j)_m=2}
\end{figure}

Fig.~\ref{wr(j)_m=2} shows the impact of rotation on the $m=2$ mode frequency for different values of $\chi$. In the case of SASI modes ($\chi=0$ and $\chi=2$), the frequency increases almost linearly with the rotation rate for $j$ varying from $0$ to $0.5$. The convective mode corresponding to $\chi=6$ has a more complex behaviour. For small rotation ($j\lesssim0.1$), the frequency is approximately proportional to the rotation rate similarly to the convective modes in Fig.~\ref{evol_wi} and \ref{evol_wi_dis}. This behaviour can be interpreted with an approximately constant corotation radius located between $r_{\rm gain}$ and $r_{\rm BV}$ (Fig.~\ref{rcorot_m2mdom}), as expected for the convective instability and visualised by the grey-shaded region in Fig.~\ref{wr(j)_m=2}. At higher rotation rates $j>0.2$, the frequency separates from this linear trend, corresponding to the corotation radius moving to larger radii (Fig.~\ref{rcorot_m2mdom}) when the mode takes a mixed nature. At still higher rotation frequencies ($j\gtrsim 0.3$), the frequencies expressed in units of $v_{\rm sh}/r_{\rm sh}$ converge toward a single linear trend that does not depend on the convection parameter. The effect of heating has been erased by rotation. This is consistent with the behaviour of the dominant modes seen earlier.

Assuming a PNS radius of $50$~km, the inverse advection timescale without heating nor rotation is $v_{\rm sh}/r_{\rm sh}\sim30$~Hz. With increasing rotation rates, this frequency decreases to $\sim25$~Hz when $j=0.5$. As a result, in Fig~\ref{wr(j)_m=2}, the frequency of the mode $m=2$, without heating, varies from $\sim120$~Hz to $250$~Hz. In the case $\chi=6$, the frequency of the $m=2$ mode is in the range [0,80]~Hz. These frequencies overlap with the optimal detection frequency range of the current gravitational wave detectors. 

%----m=1*2 vs m=2

\begin{figure}[h]
    \centering
    \includegraphics[width=\hsize]{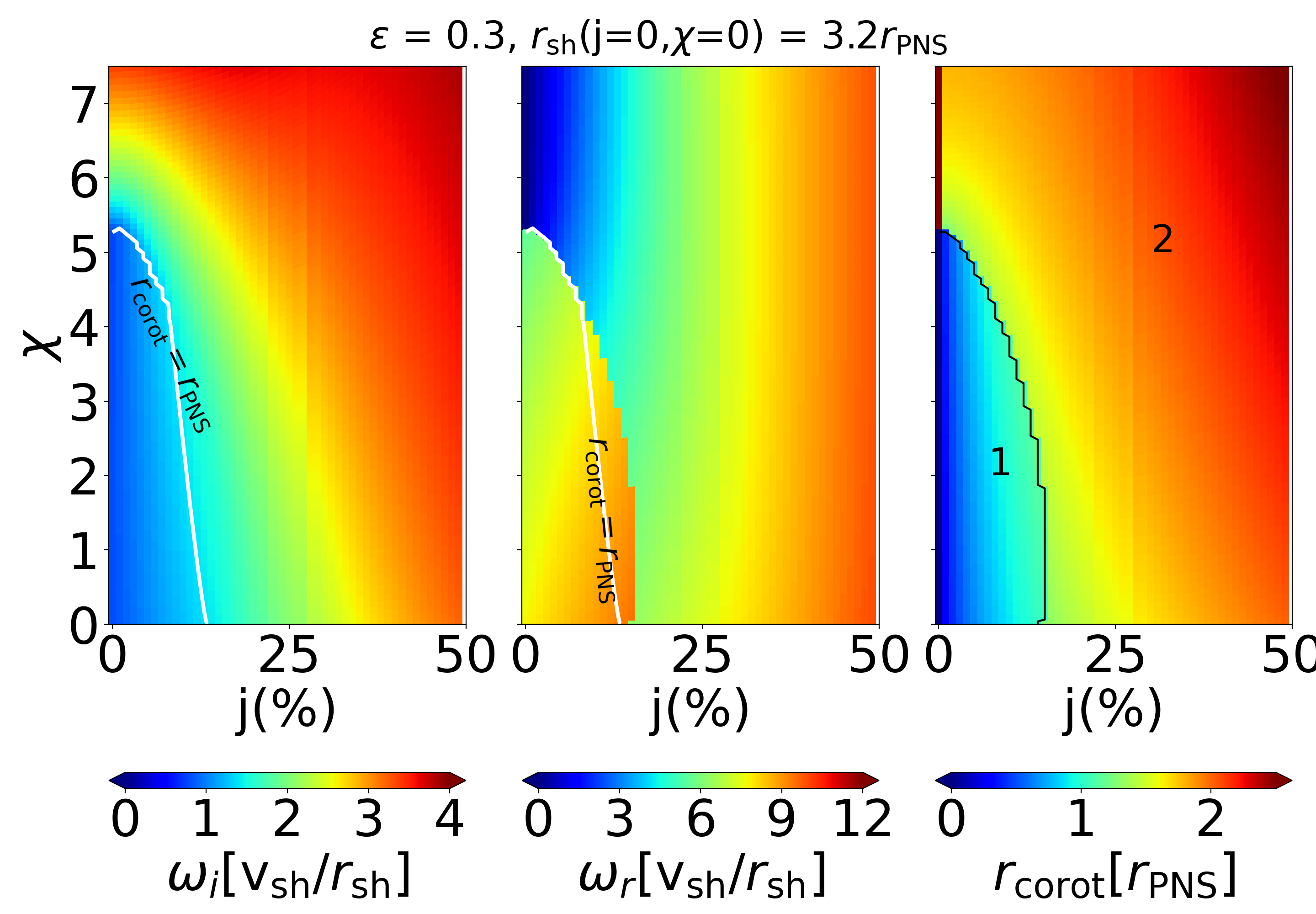}
    \caption{Maps, for varying heating and rotation rates, of the growth rate, the frequency and the corotation radius of the most unstable fundamental mode between modes $m=1$ and $m=2$. In the second panel, we plotted $2\omega_r$ when $m=1$ is dominant to account for the doubled frequency expected in the gravitational wave signal. In the third panel, the black line outlines the frontier between the dominant mode $m=1$ or $m=2$.}
    \label{plan_m=2}
\end{figure}
To account for the possible signature of a dominant mode $m=1$ on the gravitational wave signal, we include in Fig.~\ref{plan_m=2} the doubled frequency of the mode $m=1$ when it is more unstable than the mode $m=2$. Note that this frequency is higher than the frequency of the $m=2$ mode. This remarkable property can be deduced from Figs. 5 and 6 in \cite{Foglizzo+2007} where the frequency of the SASI mode $m=2$ is always smaller than twice the frequency of the mode $m=1$. We observe that the region of the parameter space dominated by the mode $m=1$ is limited to moderate convection parameters $\chi<5.2$ and modest rotation rates $j<15\%$.

In the regime $j\ge30\%$, an analytical form of the linear dependence illustrated by Fig.~\ref{wr(j)_m=2} would be useful to directly translate the frequency observed in the gravitational wave signal into a relation between $j$ and the advection frequency at the shock $v_{\rm sh}/r_{\rm sh}$.

\section{Conclusion and discussion}

In this paper, we studied a spherical-equatorial toy model where we considered the dynamics of the equatorial plane between the shock and the surface of the PNS. Thanks to a linear analysis, we computed unstable non-axisymmetric modes for varying rotation and heating rates and different dissociation energies through the shock. This led to several discoveries concerning the properties of the convective instability with rotation and its interplay with SASI and rotation-induced instabilities.

Relatively slow rotation rates (less than $10\%$ of the Keplerian rotation at the PNS radius), for which centrifugal effects are very small, can have a strong and complex influence on both SASI and neutrino-driven convection:
\begin{itemize}
    \item In the presence of rotation, convective modes have an oscillation frequency that corresponds to a corotation radius located close to the maximum of the Brunt-V\"ais\"al\"a frequency, where the convective engine is expected to be most powerful. The appearance of this non-vanishing oscillation frequency blurs somewhat the abrupt transition between SASI and convection.

    \item For large values of the convection parameter $\chi \gtrsim 5$, rotation hampers the growth of non-axisymmetric convective modes. The decrease in growth rate is more pronounced for larger $m$ modes, which can be interpreted by the larger stabilising effect of the shear due to differential rotation. In this regime, axisymmetric modes are expected to dominate the dynamics, because they are only slightly affected by rotation. 
    
    \item By contrast, for $\chi\lesssim 5$, where the advection has a strong influence on the growth of convection, rotation increases the growth rate of both SASI and the convective instability. The increase in the SASI growth rate with rotation was well known in the absence of heating \citep{YamasakiFoglizzo2008,Kazeroni+2017,Blondin+2017} and is here shown to hold in the presence of neutrino heating. On the other hand, this behaviour was not expected for the convective instability, which was generally thought to be hampered by rotation. The destabilising effect of rotation on convection in this regime leads to a decrease of the value of $\chi$ defining marginal stability. We interpret this effect as a residual influence of an advective-acoustic cycle which acts to reinforce convective motions close to the transition between convection and SASI. 

    \item We observe two different effects of rotation on the dominant scale, depending on the instability mechanism that dominates without rotation. For SASI, the dominant $m$ increases with rotation (as in \citealp{YamasakiFoglizzo2008} and \citealp{Blondin+2017}), while it decreases for the convective instability.
\end{itemize}

At moderate rotation rates ($10$ to $30\%$ of the Keplerian rotation at the PNS surface), the modes are so modified by rotation that it becomes impossible to distinguish clearly between convective and SASI modes. When the heating rate is increased, the frequency and azimuthal number of the most unstable mode change smoothly, with no clear transition from one instability to another. Modes retain characteristics of both instabilities and should therefore be understood as mixed SASI/convection/rotation modes. The dominant mixed modes have a relatively large angular scale with $m=1-3$ and their growth rate increases with faster rotation. This regime takes place above a critical angular momentum such that the frequency of a convective mode (corotating with the radius of maximum Brunt-V\"ais\"al\"a frequency) is comparable to the frequency of SASI. The non-oscillatory nature of convection and the low frequency associated to SASI both allow for the appearance of a corotation radius at moderate rotation rates. The corotation radius enables the extraction of energy and angular momentum from the interior region rotating faster than the region exterior to the corotation \citep{Cairns1979,YoshidaSaijo2017,SaijoYoshida2006}. For a PNS radius of $r_{\rm PNS}=50$~km, the mixed mode regime corresponds to an interval of specific angular momentum $3-9\times 10^{15}$~cm$^2$/s. Extrapolated to the radius $\sim 12\,{\rm km}$ of a cold neutron star, this corresponds to relatively fast rotation periods of $\sim 1 - 3 $ms. %12km de rayon (2*pi/omega, omega=J/r^2

%Regime des rotations rapides quasiment indépendant du chauffage
For high rotation rates (more than $30\%$ of the Keplerian frequency at the PNS surface, i.e. $j\gtrsim 0.3$), the frequency and growth rate are independent or weakly dependent on the heating rate, when they are expressed in terms of the advection timescale $r_{\rm sh}/v_{\rm sh}$. Together with the significant deviation of the corotation radius from the most buoyant region, this suggests that the instability is dominated by rotational rather than buoyancy effects. The study of
\cite{Walk+2022} without neutrino heating pointed out the existence of a similar instability regime where the frequency of the dominant mode depends too little on the advection time to be explained by an advective-acoustic cycle. Our results suggest that these results obtained without heating are still valid when heating is taken into account. We note that the regime of rapid rotation appears for $j \sim 0.3$ which corresponds to a small ratio $\sim 0.03$ of the centrifugal force to the gravity at the corotation radius, even though the centrifugal displacement of the stationary shock is not negligible. A precise understanding of this strong rotation regime is still missing, but the corotation radius seems to play an important role in the instability mechanism. 

%Résumé des conséquences pour les ondes gravitationnelles.
A future detection of gravitational waves is expected to give information on the physical phenomena during stellar core-collapse. 
Our identification of three different instability regimes depending on the rotation rate should help clarify the still poorly known influence of rotation on the gravitational wave signal. Non-axisymmetric convective modes become oscillatory in the presence of rotation. Their frequency could be identified in the low-frequency part of the gravitational wave spectrum. For low/moderate rotations, the corotation radius of the $m=2$ convective/mixed mode is close to the gain or Brunt-V\"ais\"al\"a radius. The identification of the mode frequency would therefore give access to the rotation frequency at this radius.
Non-axisymmetric equatorial modes with a large angular scale $m=1,2$ are strongly destabilised by rotation and dominate the dynamics in a wide region of the parameter space for moderate to strong rotation, which should be favourable to a strong emission of gravitational waves.
In the regime of strong rotation, the frequency of the $m=2$ mode becomes independent of the convection parameter and depends only on the advection timescale $r_{\rm sh}/v_{\rm sh}$ and the angular momentum $j$. This reduction of the parameter space should help extract physical information from the measure of the mode frequency. These modes should be incorporated in future asteroseismic studies similar to \cite{TorresForne+2018,TorresForne+2019} where non-axisymmetric perturbations would be taken into account.

Our results can be compared to previous 3D numerical simulations of core-collapse supernovae, including rotation. Simulations with the fastest rotating rates are difficult to compare because they are often dominated by the low-T/|W| instability growing rapidly inside the PNS \citep{Ott+2005,CerdaDuran+2007,kuroda+2014,Takiwaki+2016,Shibagaki+2021,Takiwaki+2021,Bugli+2022}, which is not included in our analysis. We focus our discussion on the simulations where the dynamics is dominated by postshock instabilities. Some simulations focused on the GW signal \citep{westernacher-schneider19,PowellMuller2020,Pan+2021} do not contain a detailed enough description of the post shock dynamics to allow for a significant comparison.

% Comparaison Fryer
\cite{FryerHeger2000} and \cite{FryerWarren2004} found that convection was quenched by rotation, especially in the equatorial plane. For models A and B of \cite{FryerWarren2004}, this may be explained by the strong centrifugal effects for such strong rotation ($j^2>0.4$, Eq.~\ref{eq:forces}). Convection in the slow rotating model C ($j\sim0.03$) does not seem to be confined to the poles, which is consistent with our study, where a purely convective instability can be observed in the equatorial plane up to $j\sim0.1$. %j = \Omega(a 1000km)*0.5/1.5

%Comparaison Iwakami
The setup of \cite{Iwakami+2014} is similar to ours in that it did not involve the interior of the PNS and considered a similar range of rotation rates (corresponding to $j=0-0.33$ in our units). In models D, E, and F as displayed in Fig.~5 of their study, for higher rotation and smaller neutrino luminosity, the scale of entropy structures is larger and buoyant patterns become spiral ones. This structure change is consistent with the rotation-induced transformation of convective modes into mixed spiral modes with a larger angular scale. The precise interpretation of their results is, however, complicated by the fact that both rotation and the neutrino luminosity are changed such that it is difficult to disentangle their respective influence. 
In addition, we note that they identified a pattern referred to as a spiral motion with buoyant-bubble (SPB), which  may be related to the mixed SASI/convection/rotation modes identified for moderate rotation rates. 

%comparaison summa
The 3D simulations of \cite{Summa+2018} show two models with specific angular momenta $\sim 5\times10^{14}$~$\rm cm^2/s$ and $\sim 10^{16}$~$\rm cm^2/s$ at the shock ($\sim 150$~km) during its stalled phase ($\sim$150~ms). These values correspond to $j\sim 1\%$ and $j\sim33\%$, respectively. In the low rotation case, convection is observed in the stalled phase of the shock. This behaviour is consistent with our results showing that a convective behaviour is possible for $j<10\%$. For the fast rotating simulation, the large scale ($l=1$) instability associated to SASI, without a tightly wound spiral pattern, is consistent with the morphology of our eigenmode in the fast rotating regime (Fig. \ref{structure}). Although Fig.~\ref{pano} suggests that the fastest growing mode corresponds to $m=2$ rather than $m=1$ in this regime, the Fig.~12 in \cite{Summa+2018} showing a dominant $l=2$ component in the turbulent energy spectrum calls for a more detailed comparison of the $m=1, 2$ components in the simulation.

To gain a better understanding of the processes at work at the onset of the shock revival, a non-linear analysis of the dynamics of the fluid would be necessary. It would shed light on how rotation acts on the saturation mechanism \citep{guilet10b} and show how the system evolves from the linear phase to the explosion. This last part would give information on the signatures of linear phenomena that may remain observable in the multimessenger signal.

One should keep in mind that our model is idealised in many respects. We assumed an ideal gas equation of state with $\gamma=4/3$, dissociation was taken into account as a fixed energy sink, neutrinos were parameterised through analytical cooling and heating terms. We also did not describe the equatorial swelling of the PNS due to the centrifugal force. This effect increases the neutrinosphere radius, leading to cooler neutrinos and a smaller heating rate in the equatorial plane. As discussed above, such a centrifugal effect is expected to be very small at low/intermediate rotation rates ($j<0.3$) but can become significant for the fastest rotations considered in this analysis. Part of this complexity is avoided by displaying our results as a function of $\chi$ rather than the heating rate. Note that when $\chi$ is kept constant and the rotation increased, the heating constant $\tilde{A}_{\rm h}$ decreases, in a qualitatively similar way to the expected impact of the neutrinosphere swelling.

The centrifugal force induces a deformation with respect to the spherical symmetry \citep{fujisawa19}, which is challenging to take into account in our formalism because it would couple different spherical harmonics. To avoid such a complexity, we restrained our analysis to the equatorial plane while keeping the effect of radial convergence in spherical geometry. As a result, our conclusions are limited to modes with spherical harmonics indices $m=\pm l$ and cannot describe the dynamics outside the equatorial plane, such as convective motions along the polar axis. The equatorial non-axisymmetric modes described here are nonetheless expected to dominate the dynamics in most of the parameter space except at $\chi\gtrsim5 - 5.5$ and slow to moderate rotation, where axisymmetric convective modes are expected to be more unstable. 

The highest specific angular momentum considered in this study should lead to the development of the low-$T/|W|$ instability inside the PNS \citep{Takiwaki+2021,Bugli+2022}. Being restricted to the post-shock region without including the PNS interior, our linear analysis cannot describe the low-$T/|W|$ instability and may miss the most unstable mode in the regime of strong rotation. A linear stability analysis including both the post-shock region and the PNS interior is therefore an important next step. Previous linear mode calculations focused on the prediction of GW mode frequencies and included the post-shock region in addition to the PNS interior, but they were restricted to axisymmetric modes, and they assumed a hydrostatic equilibrium neglecting advection \citep{TorresForne+2018,TorresForne+2019}. To describe all unstable mode and their possible interaction, a linear stability analysis will have to face the challenge of combining the PNS in approximate hydrostatic equilibrium and the advection in the post-shock region.  

Finally, the magnetic field neglected in this study is expected to play an important role for strong rotation. A strong magnetic field can for example quench the development of the low-$T/|W|$ instability \citep{Bugli+2022}. Even in the absence of rotation, the magnetic field can have a complex influence on the  post-shock dynamics because of the propagation of vorticity through Alfvén waves \citep{Guilet10a,Guilet11} and a small-scale dynamo can amplify its strength \citep{Endeve+2012,mueller20_magnetic}. 
The magnetic field should thus be taken into account in future studies of post-shock instabilities.

\section*{Acknowledgements}
JG acknowledges support from the European Research Council (MagBURST grant 715368). EA is supported by RK MES grant No. AP13067834 and NU Faculty Development Grant No. 11022021FD2912. 


%\bibliographystyle{aa}
\bibliography{biblio}

% \begin{thebibliography}{}

\end{document}